\documentclass[useAMS,usenatbib,fleqn]{mn2e}

\usepackage{times}
\usepackage[utf8]{inputenc}
\usepackage[T1]{fontenc}

\usepackage{natbib}
\usepackage[pdftex]{graphicx}
\usepackage{amsmath}
\usepackage{siunitx}

\usepackage{ltxtable}
\usepackage{supertabular}
\usepackage{array}
\usepackage{booktabs}

\usepackage{aas_macros}

\usepackage{hyperref}

\usepackage{flafter}

\input{mathdefs.sty}
\newcommand{\usk}{\,}
\newcommand{\Msun}{M_{\odot}}
\newcommand{\Medd}{\dot{M}_{\text{Edd}}}
\newcommand{\mdot}{\dot{m}}

\newcommand{\rv}[1]{#1}
\newcommand{\mathrv}[1]{#1}

\newcommand{\vrel}{v_\text{r}}
\newcommand{\vkep}{v_\text{K}}
\newcommand{\Porb}{P_\text{orb}}
\newcommand{\incl}{\theta_\text{i}} 
\newcommand{\incp}{\delta} 
\newcommand{\tdyn}{t_{\text{dyn}}}
\newcommand{\thyd}{t_{\text{h}}}
\newcommand{\tdel}{t_\text{delay}}
\newcommand{\tdiff}{t_{\text{d}}}

\newcommand{\toutb}{t_{\text{outb}}}
\newcommand{\ffcutoff}{\psi_\text{ff}}
\newcommand{\Ledd}{L_{\text{Edd}}}
\newcommand{\Teff}{T_{\text{eff}}}

\newcommand{\Lmax}{L_{\text{max}}}
\newcommand{\Lthin}{L_\text{t}}
\newcommand{\Ltot}{L_\text{tot}}
\newcommand{\Ldiff}{L_\text{d}}
\newcommand{\epsff}{\epsilon^{\text{ff}}}
\newcommand{\rschp}{r_{S,1}}

\newcommand{\rbh}{R_\text{BH}}
\newcommand{\taueff}{\tau_\text{e}}
\newcommand{\taua}{\tau_\text{a}}
\newcommand{\taus}{\tau_{\sigma}}
\newcommand{\Tvir}{T_\text{vir}}
\newcommand{\Teq}{T_\text{eq}}
\newcommand{\Tdisc}{T_\text{disc}}
\newcommand{\expfac}{\xi}
\newcommand{\expfacff}{\expfac_\text{ff}}
\newcommand{\expfacLV}{\expfac_\text{LV}}
\newcommand{\expfacT}{\expfac_\text{T}}
\newcommand{\thickratio}{\lambda}
\newcommand{\mach}{\mathcal{M}}

\newcommand{\Prad}{P_\text{r}}
\newcommand{\Pgas}{P_\text{g}}
\newcommand{\dilution}{\epsilon}

\newcommand{\ionfrac}{X}

\newcommand{\pn}{n_{14}}
\newcommand{\ph}{h_{15}}
\newcommand{\pmp}{M_{10}}
\newcommand{\pms}{M_{8}}
\newcommand{\pr}{r_{10}}
\newcommand{\lumdist}{D_L}
\newcommand{\plumdist}{D_{L,1500}}

\newcommand{\hel}{1}
\newcommand{\tuo}{2}

\title[]{Black hole accretion disc impacts}
\author[P. Pihajoki]{%
    P. Pihajoki$^{\hel,\tuo}$\thanks{E-mail: pauli.pihajoki@iki.fi} \\
    \\
    $^{\hel}$ Department of Physics, University of Helsinki, Gustaf
    Hällströmin katu 2a, FI-00560 Helsinki, Finland \\
    $^{\tuo}$ Tuorla Observatory, Department of Physics and Astronomy, University of Turku,
    Väisäläntie 20, FI-21500 Piikkiö, Finland \\
}

\begin{document}

\date{}

\pagerange{\pageref{firstpage}--\pageref{lastpage}} \pubyear{2015}

\maketitle

\label{firstpage}

\begin{abstract}
We present an analytic model for computing the luminosity and spectral
evolution of flares caused by a supermassive black hole impacting the
accretion disc of another supermassive black hole.
Our model includes photon diffusion, emission from optically thin
regions and relativistic corrections to the observed spectrum and
time-scales.
We test the observability of the impact scenario with a
simulated population of quasars hosting supermassive black hole binaries.
The results indicate that for a moderate binary mass ratio of $0.3$, and impact
distances of $100$ primary Schwarzschild radii,
the accretion disc impacts can be expected to equal or exceed
the host quasar in brightness at observed wavelength $\lambda=510\,\si{nm}$ up to $z=0.6$.
We conclude that accretion disc impacts may function as an independent probe for
supermassive black hole binaries. We release the code used for computing the
model light curves to the community.
\end{abstract}


%

\section{Introduction}\label{sc:intro}

Several studies have investigated the effect of an object travelling
through or impacting an accretion disc. Most of these studies have focused on
the interaction of stars or black holes with the accretion disc of a
supermassive black hole (SMBH) in the centre of a galaxy. Two main lines of
study can be identified in this context. One is the effect of the impactors on the disc
dynamics, such as angular momentum transport \citep{ostriker1983}, mass
deposition \citep{armitage1996,miralda2005}, disc heating
\citep{perry1993,mckernan2011},
disturbances or truncations of the disc \citep{lin1990,macfadyen2008} and
modulations of the disc-SMBH accretion rate \citep{mckernan2011}.
A second main focus has been the effect of the existing disc or disc formation on the stellar
population near the central SMBH
\citep[e.g.][]{vokrouhlicky1998a,vokrouhlicky1998b,alig2011}
or a binary SMBH embedded in the disc \citep[e.g.][]{shi2012,aly2015}.

The observable characteristics of these impact events have been
studied to a lesser extent.
On smaller spatial scales,
\citet{landry1998} presented an analytic calculation of cometary impacts
on neutron star accretion discs, and found that explosively expanding
outflows of radiation dominated gas should result, as well as bursts of
high-energy radiation.
\citet{zentsova1983} considered a star impacting the accretion disc of a
SMBH, and found that the event should leave a radiating hotspot on the surface
of the disc, but did not consider gas flowing out from the impact site.
In a seminal paper,
\citet{lehto1996} (LV96, from here on) built on this result to investigate
the impact of a SMBH on the accretion disc of another SMBH. LV96 used this mechanism to explain
the recurrent optical outbursts of the blazar OJ287, a
candidate host for a supermassive binary black hole
\citep{sil1988,val2008,val2011a}.
They found that the shocked gas should form a radiation pressure
dominated outflow, which would radiate prodigiously after turning
optically thin, but did not give specific estimates of the
resulting light curves or spectral evolution apart from simple scaling
laws. \citet{ivanov1998} (I98 hereafter) investigated the same scenario as LV96,
presenting analytic hydrodynamical results as well as a simple
2-dimensional hydrodynamical simulation. Their results mostly
agree with LV96 apart from the predicted luminosity, which was based on
analytic estimates for type II supernova light curves by
\citet{arnett1980} (A80 hereafter).
We will review both estimates and show that the slight differences in the
obtained luminosity and its evolution can be explained
by differing assumptions of the outburst mechanism, and the particular
interpretation of the A80 results in I98.
\citet{nayakshin2004} predicted X-ray
bursts from the impact of stars on the accretion disc of the
SMBH in the centre of our Milky Way galaxy. The analytic estimates for the
gas shocked by the impact are broadly in line with LV96 and I98, but no
explicitly time-dependent light curves are provided. Finally,
\citet{karas1994} and \citet{dai2010} studied the relativistic complications of
calculating light curves for impacts of stars with the accretion disc of a SMBH.
Specifically, both studies investigated the gravitational lensing, Doppler
boosting and relativistic time lag effects on the light curve.
These were calculated by combining relativistic models of the orbits with a
ray-tracing approach.
However, in both approaches the model of the impact light
curve used was simplistic, and no absolute flux estimates were given. As
such, the models are mostly useful for analyzing the relative timing of
recurrent impact outbursts.

In this work, we will extend this previous analytic work on
accretion disc impacts, with a focus on SMBH impacts on SMBH accretion discs with
mildly relativistic relative velocities.
Specifically, we will review the estimates of
in LV96 and I98 and elaborate on some of the more opaque results in LV96.
For our model, we use these results in combination with
the analytic approach presented in A80 to estimate the spectral
evolution of the outburst.
Our primary result is the estimate
of the outburst spectral flux and its time evolution. This result can in
principle be combined with the results of e.g. \citet{karas1994} and
\citet{dai2010} to obtain realistic long term light curves of repeated accretion
disc impacts.

The results of our model indicate that accretion disc impacts from SMBH binaries
in active galactic nuclei (AGN) are readily observable in IR-UV
wavelengths at cosmological distances.
The disc impact outbursts present a characteristic spectral evolution, which can be
used to identify these events. If found, recurrent outbursts could be
used to probe the parameters of the central supermassive black hole, as
in the case of OJ287 \citep[][LV96]{sil1988}. As such, these events would
provide a valuable and independent tool in estimating the masses and
spins of supermassive black holes in nearby AGN. Accretion disc
impacts might even be used to identify new SMBH binaries both by the
spectral evolution and the fact that the outbursts will have an
identifiable pseudoperiodicity \citep{karas1994}.

Our paper is organized in the following manner. In
Section~\ref{sc:disc_impact}, we will present basic analytic results of
the impact shock and its subsequent evolution. In
Section~\ref{sc:model}, we extend the analytic approach used in A80 to
construct a new outburst model.  In Section~\ref{sc:observational_tool},
we test the observability of the impacts at cosmological distance
using simulated data constructed from observed quasar mass and
luminosity distributions.  Finally, in Section~\ref{sc:discussion} we
discuss the disc impact problem, the various assumptions used to make it
analytically tractable, and the validity of these assumptions. We finish
in Section~\ref{sc:conclusions} with our conclusions.

\section{Disk impact and subsequent evolution}\label{sc:disc_impact}

We will now briefly review some elementary analytic results for a collision
between a black hole and an accretion disc. We will study
and compare the different estimates in LV96, I98 and \citet{nayakshin2004}.
All these papers consider impacts on the accretion disc of a SMBH, but with
important differences: LV96 study a SMBH impacting the inner region of a thin
disc, which is radiation dominated. I98 consider similar impacts, but in the gas
pressure dominated region. \citet{nayakshin2004} focus on impacts of stars on
the cool outer regions of the disc.
After the brief review, we will revisit the calculations of LV96 in detail for
two principal reasons. Firstly, they focus on the case most interesting
to us.
Secondly, the original paper is somewhat opaque, since only the final
results are given.

Our focus is impacts happening at mildly relativistic
velocities, i.e. $[1-(v/c)^2]^{-1/2} \lesssim 2$, near the inner parts of the
accretion disc of a supermassive black hole, around
$10\text{--}100$ Schwarzschild radii from the centre of
the hole. Of primary interest are impactors that orbit the central black hole,
since in this case the impacts and outbursts will be recurrent,
although pseudoperiodic \citep{karas1994}.
As such, while the analysis works for single transits as well, we
will refer to the impactor as the secondary, and the
accretion disc host as the primary.
Furthermore, while we consider the case of a black
hole as the impactor, the analysis does extend to stars and 
other compact objects as well.

To normalize our results,
we use $\pmp =
M_1/(10^{10}\Msun)$ and $\pms = M_2/(10^8\Msun)$ to parametrize the primary and
secondary black hole masses, respectively, and $\pr =
r/(10\rschp)$ for the impact distance,
where $\rschp = 2.95\times 10^{15} \pmp\,\si{cm}$ is the primary black hole
Schwarzschild radius.
In addition, we use $\pn = n/(10^{14}\,\si{cm})$
and $\ph = h/(10^{15}\,\si{cm})$ to parametrize the accretion disc number
density and semiheight at the impact site, while otherwise leaving the
accretion disc model unspecified. The luminosity distance is
parametrized by $\plumdist = \lumdist/1500\,\si{Mpc}$, where $\lumdist$ is the luminosity distance in megaparsecs.
The value $\plumdist=1$ corresponds to $z=0.281$
assuming a cosmology with $H_0=67.11\usk\text{km/s}/\text{Mpc}$,
$\Omega_M=0.3175$ and $\Omega_{\Lambda}=0.6825$
\citep{planck2013}.
Later, we will use the term nominal values to mean setting all normalized quantities equal to one.
The normalization constants were
chosen for easy comparison with the original work in LV96, and reflect
typical values for a Sakimoto-Coroniti $\beta$-disc \citep{sakimoto1981}
with a central mass of $\pmp\sim 1$ and an accretion rate of $0.1\Medd$, where
$\Medd$ is the Eddington accretion rate.
For most disc models,
the parameters $\ph$ and $\pn$ will depend on the impact distance $\pr$, primary
mass $\pmp$ and the primary accretion rate $\dot{M}$.

For convenience, most of the symbols have been listed with explanations
in Appendix~\ref{sc:symbol_list}.

\subsection{Impact shock}

\begin{figure}
    \includegraphics[width=\columnwidth]{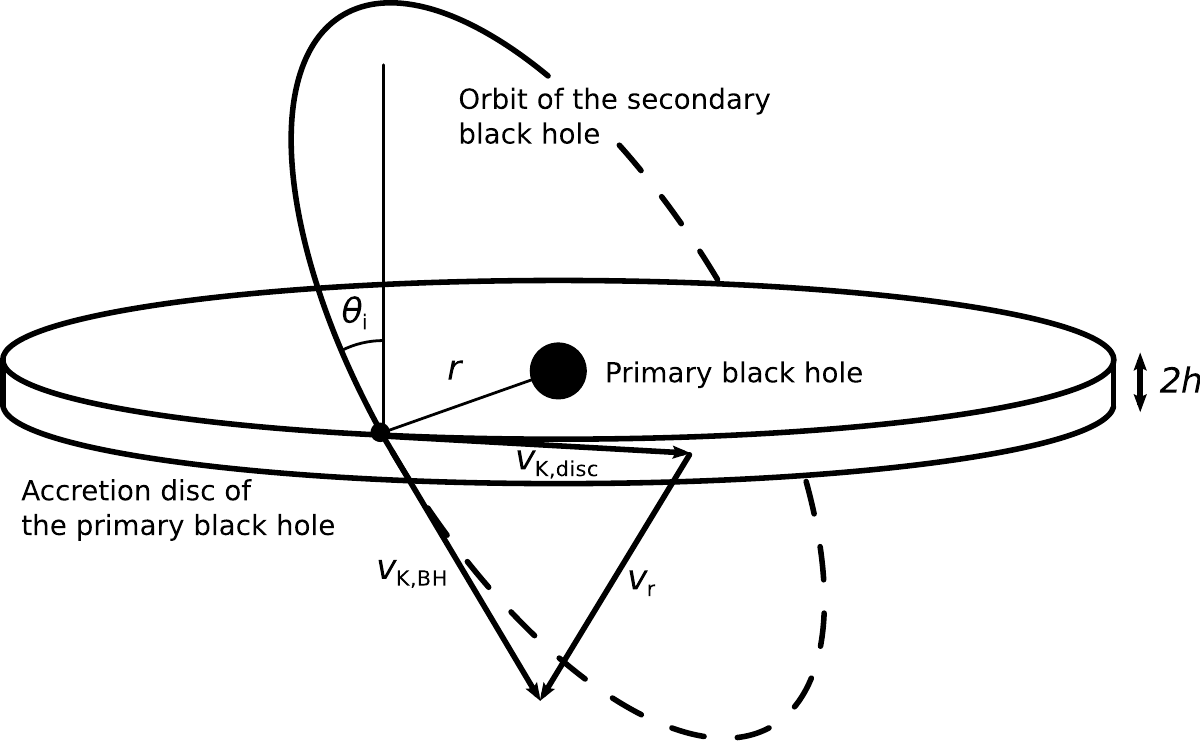}
    \caption{\label{fig:impact_geometry}
        A sketch of the accretion disc impact. The local Keplerian velocities
        of the disc matter ($v_{\text{K,disc}}$) and the secondary black
        hole ($v_{\text{K,BH}}$) are shown,
        along with the relative impact velocity $\vrel$ and the angle $\incl$.
    }
\end{figure}

Broadly speaking, the impact of the black hole with the accretion
disc of the primary happens in the following manner.
In the first stage, the secondary black hole approaches the disc and the
disc midplane is pulled towards the secondary by the gravitational
interaction.
The secondary then plunges into the disc
with a velocity $\vrel$ relative to the disc gas.
Here, we assume an orbiting
impactor with a circular Keplerian orbit, which makes an angle $\incl$ with the
normal to the disc plane. The impact geometry is illustrated in
Figure~\ref{fig:impact_geometry}.

The local Keplerian velocity is
$\vkep = \mathrv{c(2\sqrt{5})^{-1}\pr^{-1/2}}$, and the relative velocity between the
impactor and the disc is $\vrel = \sqrt{2}\incp\vkep$, where
$\incp = \sqrt{1-\sin{\incl}}$. We are mostly interested in cases
where $\incl\sim \mathrv{0}$ and $\incp\sim 1$, since an orbiting black
hole rotating close to the plane of the disc will not be a source of
impacts. Instead, it will evacuate an annulus in the disc, or cause
the disc to be truncated into a circumbinary disc \citep[see e.g.][]{artymowicz1996,macfadyen2008,farris2014}.
When $\incp \sim 1$, we have for thin $\alpha$-discs
\citep{shakura1973} that
$\vrel/c_s\sim \sqrt{2}\incp\alpha^{-1} \gg 1$, where $c_s$ is the sound
speed in the disc. As such, the impact can
be expected to be highly supersonic, and the disc gas will be strongly
shocked.

The dynamical time, $\tdyn$, of the impact
event can be estimated from the disc crossing time
\begin{equation}
    \tdyn = \frac{2h}{\vkep\cos\incl} \sim 3.0\times
    10^{5}\ph\pr^{1/2}[1+2(\incp-1)^2]\,\,\si{s}.
\end{equation}
Since the impact velocity is high, the ratio of the initial to virialized 
energy of the disc matter is
\begin{equation}
    \frac{3k\Tdisc}{m_p\vrel^2} \sim \mathrv{2.8}\times 10^{-6} \pr\incp^{-2}
    \left(\frac{\Tdisc}{10^6\,\si{K}}\right)
    \ll 1,
\end{equation}
where $\Tdisc$ is the accretion disc temperature.
We see that the internal energy of the disc gas can be ignored. If we assume
complete thermalization of the disc matter at the shock, the
temperature of the protons in the postshock region is then close to the virial
temperature
\begin{equation}
    \Tvir \sim \frac{m_p \vrel^2}{3 k} \sim \mathrv{3.6}\times 10^{11}\pr^{-1}\incp^{2}\,\si{K},
\end{equation}
where $m_p$ is the mass of the
proton. The initial temperature of electrons is smaller by a factor of
$m_p/m_e\sim 1836$, but Coulomb interactions
thermalize the electron population with the ions essentially immediately \citep{weaver1976}.
The temperature $\Tvir$ is very high, and the postshock region will
be dominated by radiation pressure, up to impact distances of
$\pr\sim 1\times10^7 \pn^{-1/3}\incp^2$.
The postshock number density $n_2$ is then given in
terms of the number density of the disc matter, $n$, by
\begin{equation}
    n_2 = \frac{\gamma_a+1}{\gamma_a-1}n = 7n,
\end{equation}
where $\gamma_a=4/3$ is the adiabatic constant
for a radiation dominated mixture and the shock compression
ratio is equal to $7$.

\rv{The shocked matter can cool via several mechanisms.}
The bremsstrahlung cooling time-scale $t_\text{ff} = u/\epsff$,
or in other words, the ratio of the plasma energy density $u$ to the
\rv{(relativistic) hydrogen bremsstrahlung
emissivity $\epsff$ is \citep[e.g.][]{bethe1934,longair1997}
\begin{equation}\label{eq:fftime-scale}
    \mathrv{
t_\text{ff} = 
\left\{ 4\alpha c r_e^2 Z^2 n_2\left[\log\left(\frac{183}{Z^{1/3}}\right)+\frac{1}{18}\right] \right\}^{-1}
\sim 34.9\,\pn^{-1}\,\si{s},}
\end{equation}
where $\alpha$ is the fine structure constant, $r_e\approx
\SI{2.82e-13}{cm}$ is the classical electron radius, and we take $Z=1$
for hydrogen.}
The inverse Compton cooling time-scale is determined by the radiation
field already in place, which we take to originate from the disc. We get
\begin{equation}
    t_\text{IC} = \frac{u}{\epsilon_\text{IC}} =
    \frac{n_2 m_p \vrel^2/2}{\frac{4}{3}c\sigma_T a \Tdisc^4 \gamma^2\beta^2}
    \sim 2\times 10^{-3}\left(\frac{\Tdisc}{10^6\,\si{K}}\right)^{-4}\, \si{s}
\end{equation}
where $\sigma_T\approx 6.65\times10^{\mathrv{-25}}\,\si{cm^2}$ is the Thomson
electron scattering cross section,
$\gamma = 1 + 3k\Tvir/(2m_e c^2) \sim 1 + \mathrv{92}\pr^{-1}\incp^2$ is the electron Lorentz
factor, $\beta = 1-\gamma^{-2}\sim 1$ is the electron beta,
and $a = 4\sigma_T/c$ is the radiation constant.
\rv{The high energy photons produced by the cooling matter pair produce
and lose energy with a time-scale of \citep{longair1997}}
\begin{equation}
    \mathrv{
        \begin{split}
            t_\text{pp}   & \sim \left\{0.37 \alpha c r_e^2 Z^2 n_2 \left[\frac{28}{9}\log\left(\frac{2 E_\gamma}{m_e c^2}\right)-\frac{218}{27}\right]\right\}^{-1} \\
                     & \sim
                     220\left[3.1\left(4.8-\log\pr+2\log\incp\right)-8.1\right]^{-1}\,\pn^{-1}\,\si{s},
    \end{split}
}
\end{equation}
\rv{
where we have taken the photon energy to be 
$E_\gamma\sim k\Tvir \sim \mathrv{31}\pr^{-1}\incp^2\,\si{MeV}$, and the
factor $0.37$ is from assuming a thermal distribution of photons at
$\Tvir$. Pair production from photon-photon interactions is negligible.
Typically all cooling timescales are much less than $\tdyn$ and
the shocked matter-radiation mixture is thermalized when the
impact event ends (i.e. after $\sim\tdyn$).
}

\rv{The radiation is impeded from escaping the impact site due to the
high optical thicknesses involved.}
The electron scattering optical depth
orthogonal to the disc plane is
$\tau \sim h n_2 \sigma_T\sim 4.7\times10^{5} \ph \pn$, which depends on
the accretion disc model, but for the inner parts of a thin
$\alpha$-disc of a SMBH, $\tau$ is typically much greater
than one. \rv{A small fraction of the initial radiation at energies
$\gtrsim k\Tvir$ is in the decreasing Klein--Nishina tail of the
scattering cross section and may escape. By geometry, a fraction
$\sim\tau^{-1}$ of the radiation can also escape, originating from the
the outer layers of the impact site. The escaping photons
may produce a brief high energy transient with a time-scale of
$\sim\tau^{-1}\tdyn$, but since we are interested in the long time-scale
brightness evolution, we will not further investigate this transient in the
paper.}
\rv{Most} of the radiation cannot escape the optically thick
postshock region, and the matter and trapped radiation are brought into
equilibrium. The equilibrium temperature $\Teq$ of the mixture can be obtained from the
solution to a strong radiative shock \citep{pai1997},
\begin{equation}\label{eq:Teq}
    \begin{split}
        \Teq &= \left[1+\frac{8}{7}\left(\mach_{e,1}^2-1\right)\right]^{1/4}\hspace{-1em}\Tdisc
        \sim \left(\frac{8}{7}\right)^{1/4}\mach_{e,1}^{1/2} \Tdisc \\
        &=\left(\frac{18nm_p \vrel^2}{7a}\right)^{1/4}
        \sim \mathrv{1.50}\times10^{6} \pn^{1/4} \pr^{-1/4}\incp^{1/2}\,\si{K},
    \end{split}
\end{equation}
where $\mach_{e,1}=\sqrt{3\gamma_a\mach_1^2/(4R_P)}$ is the effective Mach
number before the shock, $\mach_1=\vrel/c_{s,1}$ is the Mach number and
$R_P = \Prad/\Pgas$ is the ratio of radiation and gas pressures.
From equation \eqref{eq:Teq} we see that the impact process leaves the mixture
of gas and radiation in a a temperature that depends only weakly on impactor
inclination, impact distance
and initial density of the accretion disc at the impact site.

\subsection{Outflow}

In LV96, it is estimated that the shocked gas emerges from the
disc after a delay of $\tdel=h/v_g$, where $v_g = (6/7)\vrel$ is the velocity
of the accretion disc gas perpendicular to the disc plane after the
shock.
However, the hydrodynamical simulations of I98
indicate instead that the gas outflow initiates approximately when the
secondary contacts the disc. The reality is likely somewhere between
these two extremes. The reasoning is that I98 models the disc with a top-hat density
distribution, but in a more realistic thin disc most of the mass is
close to the midplane and as such approximately one disc semiheight away
from the fiducial surface.

For analytic purposes, both models estimate
the emerging gas to form a spherical blob, with I98 estimating an
initial radius of
\begin{equation}
    R_{0}^{\text{I98}} = \rbh = \frac{2GM_2}{\vrel^2}
    \sim \mathrv{2.95}\times 10^{14} \pms \pr\incp^{-2}\,\si{cm},
\end{equation}
given by the
Bondi--Hoyle radius $\rbh$ \citep{bondi1944,bondi1952}. LV96 give an estimate
for a spherical volume, $V_0 \sim \pi \rbh^2 h/7$, derived from a
cylinder created by the Bondi--Hoyle radius and the disc semiheight,
$h$, divided by the shock compression ratio. This gives
\begin{equation}
    R_0^\text{LV96} = \left(\frac{3}{28}\rbh^2 h\right)^{1/3}
    \sim \mathrv{2.11}\times 10^{14} \ph^{1/3} \pms^{2/3} \pr^{2/3} \incp^{4/3}\,\si{cm}.
\end{equation}
The I98 simulations
seem to indicate that a better estimate for the size of the initially shocked
volume would be a cylinder formed by the disc semiheight $h$ and a
radius $R\sim h/2$, which would give
\begin{equation}
    R_0^\text{sim} = \mathrv{5.72}\times 10^{14} \ph\,\si{cm}.
\end{equation}
There is some
uncertainty in this however, since the I98 simulation
is based on a non-radiative, non-relativistic purely hydrodynamical
code, with idealized initial conditions.
Despite this, the three results are surprisingly close for the
nominal case, mostly due to the fact that $\rbh\sim h$. The essential
difference is in the dependence on the Bondi--Hoyle radius
and the disc semiheight, i.e. $R_0^\text{I98}\propto \rbh$,
$R_0^\text{LV96}\propto \rbh^{2/3} h^{1/3}$ and
$R_0^\text{sim}\propto h$. This reflects the fact that the problem is
sensitive to the ratio $\thickratio=\rbh/h = 0.30 \pms \pr
\ph^{-1}\incp^{-2}$. For
$\thickratio \ll 1$, the situation reduces to Bondi--Hoyle accretion. For
$\thickratio \gg 1$, the problem resembles a bullet-like impact on a
thin gas film. In our case, $\thickratio\sim 1$, and to find which analytic
estimate describes the size of the shocked region well would likely
require numerical simulations. It is clear however, that for cases where
$\ph < \rbh$, the estimate $R_0 = \rbh$ is unphysically large.

The simulations in I98 also show that the approximating the outflow as
spherical is surprisingly accurate for the kind of impact
considered here.
There is a slight caveat here
as well, as the I98 simulation uses a step function for the density
profile of the accretion disc. It is well known \citep[e.g.][]{sanders1976}
that a pointlike energy injection within a slab of matter with 
a density gradient causes the outflow to be more jetlike. In contrast, for
impacts in the cooler and less dense disc regions, as considered in
\citet{nayakshin2004}, no outflow has time to form.

After the initial sphere of outflow gas is formed,
the estimates in LV96 and I98 diverge. The I98 model
uses a formula for supernova luminosity from A80, given by
\begin{equation}\label{eq:I98sol}
    L(t) = \frac{\pi^2}{9}\Ledd
    \exp\left[-\left(1+\frac{t}{2\tdyn}\right)\frac{t}{\tdiff}\right],
\end{equation}
where $\Ledd$ is the Eddington luminosity of the secondary black hole,
$\tdiff\sim R_0^2\kappa_T\rho_0/c$ is the initial photon diffusion time-scale
for a homogenous sphere, and $\kappa_T$ is the Thomson electron scattering
opacity.
Arnett's formula is based on a sphere that expands freely in a
linear homologous manner, with $R(t) = R_0 + v_0t$, where $v_0$ is a constant expansion
velocity. The luminosity of the sphere is based on photon diffusion. In
deriving the above result,
I98 have set the hydrodynamical time-scale $\thyd = R_0/v_0$ equal to the disc crossing
time $\tdyn$. This implicitly sets the outflow expansion velocity to
$v_0 \sim (\rbh/h)\vrel$.
As will be shown in Section~\ref{sc:model}, the maximum luminosity $L\sim\Ledd$
of the outburst derived in I98 depends on the assumption that $R_0 = \rbh$,
which for thin discs can be an overestimate.

In the LV96 model, the sphere is taken to be homogenous, and to expand
with the (time-dependent) speed of sound. This leads to an asymptotic
behaviour $R(t)\propto t^{2/3}$, as will be shown in the following.
The sphere is initially estimated to radiate very little, and cool via
adiabatic expansion. For a photon gas, $VT^3$ is constant in an
adiabatic process, where $V$
is the gas volume, so that $T\propto R^{-1}$. The
assumption of no radiation escaping is maintained until the sphere turns
optically thin, after expanding by a factor of $\tau^{4/7}$, at which point the
luminosity is estimated by
the volume times the bremsstrahlung emissivity $\epsff$. LV96 give the
evolution of the total bremsstrahlung luminosity as $\epsff R^3\propto t^{-7/2}$.
Using the assumptions in LV96, we find instead
\begin{equation}
    L\propto \epsff R^3\propto n^2 T^{1/2} R^3 \propto R^{-7/2}\propto t^{-7/3}.
\end{equation}
To gain more insight, we now revisit the calculations of LV96 in detail.

\subsection{LV96 revisited}\label{sc:lvredux}

We now consider the derivations of the outflow evolution in LV96 in more
detail. The principal result in LV96 are the scaling relations,
equations (11)-(13), for the
observed V-band flux $S_V$, outburst duration $\toutb$ and the
delay from impact to the outburst $t_0$. We obtain the
normalization constants of the equations and present the
explicit derivation, which was omitted from LV96.

In LV96, the expansion speed of the bubble is set equal to
the speed of sound, $c_s=\sqrt{\gamma_a P/\rho}$.
For a radiation dominated
gas, the adiabatic constant $\gamma_a = 4/3$ and pressure $P=P_r=aT^4/3$ is
given by radiation pressure.
In the shock frame the postshock
velocity of the gas is $u_2=u_1/x$, where $u_1=\vrel$ is the gas infall velocity
before the shock. In the frame of the disc gas, the postshock gas
velocity is then $v_2=\vrel-u_2 = 6\vrel/7$.
This leads to $c_{s,0} = 2\sqrt{2/7}\vrel\sim \vrel$. This is of
the same order as the results of the hydrodynamical simulations in I98,
but the latter report maximal expansion velocities up to $3\vrel$.

During the adiabatic expansion, $T\propto R^{-1}$ and $P\propto R^{-4}$ for a
photon gas.
Since for gas pressure $\Pgas\propto T\rho \propto R^{-4}$,
during the expansion the ratio of radiation pressure to gas
pressure is constant. The expanding sphere stays radiation dominated
in this approximation.
In reality, increasing amount of the trapped radiation
escapes as the outer parts of the expanding sphere become optically thin.
With these assumptions, we can find the size evolution of the sphere,
\begin{equation}\label{eq:Rdiffeq}
    \dot{R} = c_s =
    \sqrt{\frac{4}{3}\frac{P_0}{\rho_0}\expfac(t)^{-1}}
    = \sqrt{\frac{4}{9}\frac{a \Teq^4}{\mu m_p n_2}\expfac(t)^{-1}} 
\end{equation}
where $\expfac(t) = R(t)/R_0$ is the expansion factor,
subscript zero indicates the initial value at $t=0$, and
$\mu$ is the mean molecular weight.
Here $R(t)$ must
represent a fiducial photospheric surface, within which the radiation stays trapped.
The solution to equation \eqref{eq:Rdiffeq} is
\begin{equation}\label{eq:Rsol}
    R(t) = R_0\left(1 + \frac{3}{2}\frac{c_{s,0}}{R_0}t\right)^{2/3} 
    = R_0\left(1 + \frac{3}{2}\frac{t}{\thyd}\right)^{2/3},
\end{equation}
where $\thyd = R_0/c_{s,0}$ is the hydrodynamical (expansion) time-scale.
The solution has the asymptotic behaviour $R\propto t^{2/3}$ reported in LV96.

LV96 next assume that the sphere will not radiate until it is optically
thin. Main contribution to opacity is from Thomson opacity, $\kappa_T =
\sigma_T/(\mu m_p)$.
From the results it is evident that an effective optical depth
$\taueff = \sqrt{\taua(\taua + \taus)}$ was used to determine when the
radiation can escape, where
$\taus \sim R\rho\kappa_T$ is the optical depth due
to the Thomson electron scattering, and
$\taua \sim R\rho\kappa_a$
is the optical depth due to absorption processes such as free-free and
free-bound absorption. An approximate form for these is given by
Kramers' law
\begin{equation}\label{eq:kramer}
    \kappa_a = 4\times10^{25}(1+X)\rho T^{-7/2}\,\si{g cm^{-2}} = K_a \rho T^{-7/2}.
\end{equation}
As such, we have
\begin{equation}
    \taueff^2 =
    K_a^2 R_0^2 \rho_0^4 T_0^{-7} \expfac^{-3}
    + K_a \kappa_T R_0^2\rho_0^3 T_0^{-7/2} \expfac^{-7/2}.
\end{equation}
The LV96 results indicate that only the latter term was kept, since in
this case we obtain from the condition $\taueff\sim\sqrt{\taua\taus}=1$ the published result
for the bubble expansion ratio at the onset of optical thinness,
\begin{equation}
    \begin{split}
        \expfacLV &= \frac{R}{R_0}=\tau_{e,0}^{4/7}=(\sqrt{\taua\taus})^{4/7} \\
                  &\sim \mathrv{44}\,\pms^{8/21}\ph^{4/21}\pn^{17/28}\pr^{53/84}\incp^{-53/42}.
    \end{split}
\end{equation}
If the first term is used, we find
\begin{equation}
    \expfacff =\tau_{e,0}^{2/3}=\taua^{2/3}
    \sim 3.1\,\pms^{4/9} \ph^{2/9}\pn^{3/4}\pr^{37/36}\incp^{-37/18}
\end{equation}
instead. While the second term decays faster with increasing $R$,
in this case it still
dominates the optical thickness, and we will follow LV96 and
use it in the following.

We can now reobtain the LV96 results for the delay between the impact
and optical thinness $t_0$, the peak $V$-band flux $S_V$ and the
fiducial outburst length $\toutb$. Recreating the results of LV96
indicates that these quantities were originally solved in the following
manner.
The delay $t_0$ between the impact and the outburst is solved from
$R(t_0)/R_0 = \expfacLV$, giving
\begin{equation}
    t_0=\frac{2}{3}\thyd(\expfacLV^{3/2}-1)\sim\frac{2}{3}\thyd\expfacLV^{3/2}.
\end{equation}
The $V$-band flux density is estimated from
\begin{equation}
    \begin{split}
        S_V &\sim \epsff_{(1+z)\nu_V}(t_0) \frac{(4/3)\pi R(t_0)^3(1+z)}{4\pi D_L^2} \\
            &\sim \epsff_{(1+z)\nu_V}(0) R_0^3\frac{1+z}{3 D_L^2}\expfacLV^{-5/2},
    \end{split}
\end{equation}
where $z$ is the redshift,
$\nu_V = c/\SI{550}{nm}$ corresponds to the Johnson--Cousins $V$-band filter,
$\epsff_\nu$ is
the bremsstrahlung volume emissivity per unit frequency and
$D_L$ is the luminosity distance.
The outburst length $\toutb$ is given by the bremsstrahlung cooling
time-scale at $t_0$,
\begin{equation}
    \toutb = \frac{aT(t_0)^4}{\epsff(t_0)} = \frac{a\Teq^4}{\epsff(0)}\expfac^{5/2}_\text{LV}.
\end{equation}
Expressing all results with our normalization, we finally obtain
\begin{gather}
    \frac{t_0}{1+z} = 1.1\times 10^7\, \ph^{13/21}\pms^{26/21}\pn^{51/56}\pr^{355/168}\incp^{-355/84}\,\si{s} \\
    \frac{\toutb}{1+z} = 5.7\times 10^8\, \ph^{10/21}\pms^{20/21}\pn^{11/28}\pr^{59/84}\incp^{-59/42}\,\si{s} \\
    \begin{split}
    \frac{S_V}{1+z} &= 0.032\,\si{mJy}\\
                    &\quad\times e^{-\ffcutoff}\ph^{11/21}\pms^{22/21}\pn^{5/14}\pr^{23/42}\incp^{-23/21}D_{L,1500}^{-2}
    \end{split} \\
    \frac{\ffcutoff}{1+z} = \mathrv{0.76}\,\ph^{4/21}\pms^{8/21}\pn^{5/14}\pr^{37/42}\incp^{-37/21},
\end{gather}
where we have taken the bremsstrahlung
gaunt factor to be $g(\nu,T)=1$. In addition we find the maximum bolometric luminosity,
\begin{equation}
    \Lmax = \epsff(t_0)V(t_0) \sim 6.1\times 10^{43}\,\ph^{1/3}\pms^{2/3}\pr^{-1/3}\incp^{2/3}\,\si{erg.s^{-1}}.
\end{equation}

We return to the issue of the use of effective optical thickness. It
should be noted that the surface $\taueff=1$ corresponds to the surface
at which the emitted photons are produced, i.e.\ the surface of last
absorption. However, when scattering dominates, or $\taus \gg \tau_a$, the
photons are seen to emanate from the surface of last scattering,
$\taus\sim 1$. As such, for luminosity estimates $\taus$
should be used, not $\taueff$. In our case, the initial ratio of optical
thicknesses is $\tau_{a,0}/\tau_{\sigma,0} = \kappa_{a,0}/\kappa_T\sim
6\times10^{-5}\,\pn^{1/8}\pr^{7/8}\incp^{-7/4}$. The ratio is proportional to
$R^{1/2}$ so the opacities will be equal after
an expansion by a factor of 
$\expfac =
(\kappa_T/\kappa_{a,0})^2 \sim 3.1\times10^8\,\pn^{-1/4}\pr^{-7/4}\incp^{7/2}$.
This should be compared with the amount of expansion required to reach
$\taus=1$, which is 
$\expfacT=\tau_{\sigma,0}^{1/2}\sim 310\,\ph^{1/6}\pms^{1/3}\pn^{1/2}\pr^{1/3}\incp^{-2/3}$,
and the expansion factor at which the outflow reaches a temperature where the
most of the hydrogen is non-ionized. If $\ionfrac$ is the limiting hydrogen ionization
fraction, we can use the Saha equation to get an estimate
\begin{equation}
    \expfac
    = -14.1\,\pn^{1/4}\pr^{-1/4}\incp^{1/2}
    \,W_{-1}\left(-2.1\times10^{-8}\,\omega^{2/3}\pn^{1/6}\pr^{1/2}\incp^{-1}\right),
\end{equation}
where $\omega = \ionfrac^2(\ionfrac-1)^{-1}$ and $W_{-1}$ is the $-1$ branch of the Lambert
$W$ function. For nominal parameter values and a limiting
hydrogen ionization fraction of $\ionfrac=0.5$, we have $\expfac\sim 300$.
Based on these estimates, it is clear that for estimating bolometric luminosity, 
Thomson opacity is the
determining factor, and the outflow does not turn optically thin until
$\expfac\gtrsim 300$. Effective optical depth does in principle determine the initial
spectrum of emitted photons, but the value of $\tau_e$ must be calculated by taking
into account the bulk motion of the plasma \citep{shibata2014}. This will be
further discussed in the Section~\ref{sc:model}.
The spectrum emitted at $\taueff=1$ is modified by Compton scattering until the
photons can escape near the surface $\taus=1$. The magnitude of this effect can be
estimated from the comptonization parameter,
\begin{equation}
    \begin{split}
        y &= \int \frac{k T_e}{m_e c^2}n\sigma_T\ud l \sim \frac{k \Teq}{m_e
    c^2}n_2\sigma_T R_0 \expfac_{LV}^{-3} \\
    &\sim 3\times 10^{-4}\,\ph^{-5/21}\pms^{-10/21}\pn^{-4/7}\pr^{-31/21}\incp^{62/21},
\end{split}
\end{equation}
where $T_e = \Teq$ is the electron temperature.
The small value indicates that in this approximation, the emerging continuum
spectrum can be well approximated by a diluted blackbody spectrum. The effective
black body temperature is then
\begin{equation}
    \begin{split}
        T_\text{eff} &=\Teq\expfacLV^{-1} \\
                     &\sim3.5\times10^4\,\ph^{-4/21}\pms^{-8/21}
\pn^{-5/14} \pr^{-37/42}\incp^{-37/21}\,\si{K},
    \end{split}
\end{equation}
and the spectral flux is multiplied by a dilution factor
\begin{equation}
    \epsilon=(\expfacLV/\expfacT)^2\sim 0.019\,\ph^{1/21}\pms^{2/21}\pn^{3/14}\pr^{25/42}\incp^{-25/21}.
\end{equation}
The dilution factor accounts for the fact that while the spectrum is formed
earlier, in less expanded, hotter outflow, it can only escape later, when the
outflow is optically thin to Thomson scattering.

\subsection{Discussion}\label{sc:discussion1}

The previous work reviewed above gives a good overview of the impact
shock. There is a clear consensus, at least for impacts with
$\thickratio \ll 1$, i.e.\ for fast impactors and thin discs.
One serious uncertainty however, is the size $V_0$ of the shocked region, or
equivalently, the magnitude of the effective cross-section $\sigma$ for the
impactor-disc interaction. Since we assume that the post-shock radiation
field is in thermal equilibrium, $V_0$ (or $\sigma$) directly scales the
energy deposited into the gas by the impactor. Unfortunately, as
discussed above, estimating $V_0$ is difficult analytically.

The details of the subsequent impact outflow, its evolution and its
observational characteristics are much less well defined.
For example,
I98 and LV96 obtain different time dependences for the
outflow photosphere radius, $R\propto t$ or $t^{2/3}$ respectively,
assuming either linear expansion or adiabatic expansion with the speed
of sound.
We find the first scenario easier to motivate physically.
Furthermore, the simulations in I98
indicate that the outflow bulk motion is ballistic, with a supersonic expansion
velocity, at least after the initial stages of the outburst.
We can compare these scenarios to results
of the point explosion problem, which has been studied extensively
since the original work by Sedov and Taylor
\citep{sedov1946,taylor1950}.
The justification is that for cases where $\thickratio \ll 1$ the impact
can be relatively quick compared to the gas dynamical time-scales.
In this case, an amount of energy
$E_0\sim V_0\rho\vrel^2$ is deposited into the shocked volume $V_0$ in
$\tdyn$, after which we expect the expansion of the shocked matter to be
at least approximately described by point explosion theory. The
situation is complicated by the fact that the impact event is not
spherically symmetric; one part of the ambient medium is thick and in a
plane (the accretion disc), and the rest is likely to be rarified and more
isotropic, but with density gradients (the accretion disc corona).

In very general terms, the expansion behaviour of baryonic matter
injected with a large amount of energy depends initially on the ratio
$\eta = E_r/(mc^2)$ of radiation energy to matter rest energy
\citep{shemi1990,piran1993,kobayashi1999}, and the density gradient of the
surrounding matter during later times. For the disc impact case,
$\eta = a\Teq^4/(n_2 m_p c^2)=0.04\,\pr^{-1}\incp^{2} < 1$, and the shocked
matter will not be accelerated to high Lorentz factors by radiative
pressure. However, the hydrodynamical simulations in I98 do indicate
that gravitational effects of the impactor do accelerate the outflow to
mildly relativistic velocities of the order of $\vrel$. The evolution
of the outflow should then initially be ballistic, with $R\propto t$.
Later it should turn over into the Sedov-Taylor expansion with
$R\propto t^{2/(5-\omega)}$, where $\omega < 5$ and $n_c(r)\propto r^{-\omega}$ 
is the number density
gradient in the surrounding medium, which in this case would consist of the
accretion disc corona, or accretion disc matter that has been blown off
the disc by previous impacts. This turnover should happen at around 
$R\sim [E_r/(m_p n_c c^2)]^{1/3}$, or when
\begin{equation}
    \expfac\sim 2\times 10^4\,\pn^{1/6}
(n_c/1\,\si{cm^{-3}})^{-1/6}\pr^{7/6}\incp^{-7/3}.
\end{equation}
Thus, with reasonable values of $n_c$, the observable outburst is likely over by
the time the turnover happens.
We conclude that the outflow should not experience large accelerations
during the observable outburst phase, and should expand approximately
linearly.

The above descriptions omit some physical processes that may affect both
the spectrum of the observable outburst and the its hydrodynamics,
namely magnetic fields, strong gravity and radiation transfer. There is
reason to expect the inner parts of an SMBH accretion disc to be
at least weakly magnetized (see Section~\ref{sc:discussion}),
which would allow the shocked matter to cool via synchrotron
radiation. Strong gravity, obviously important near $\pr\sim 1$, would
certainly influence the hydrodynamics of the outflow and break the
(hemi)spherical symmetry of the outflow.

We will not attempt to address
the first two of these complications in the following section, where we
construct our outburst model. We note that synchrotron and inverse Compton radiation from a
relativistically expanding shockwave has been studied in the literature
before, e.g.\ in \citet{blandford1977}.
We will, however, partly address the problem of
radiation transfer by taking photon diffusion and optically thin regions into
account.

Now, while the initial outflow is expected to be
optically thick, photon diffusion cannot be neglected, and the outburst
should be observable from the very beginning, unlike the description in
LV96. Furthermore, as discussed above, the outflow is optically thick to electron scattering 
until it has expanded by a factor of $\sim 300$. At this point the outflow has
cooled and rarefied to the extent that the optically thin luminosity is a fraction
of the initial diffusive luminosity.
We can estimate the magnitude of this difference by 
considering the ratio of the optically thin bremsstrahlung
emission when $\taus=1$ compared to the initial diffusive luminosity at $t=0$.
The former is given by $\Lthin = \epsff(0)\expfacT^{-13/2}$ and latter by
$\Ldiff = a \Teq^4
V_0/\tdiff$, where $\tdiff = 3 R_0^2 n_2\sigma_T/c$ is the diffusion time.
Assuming the LV96 estimate $V_0 = \pi\rbh^2 h/7$, we get
\begin{equation}
    \frac{\Lthin}{\Ldiff} \sim
    8\times10^{-5}\,\ph^{1/12}\pms^{1/6}\pn^{3/8}\pr^{25/24}\incp^{-25/12}.
\end{equation}
As such, the observed light curve
is likely to be dominated by the diffusive period. In addition, since the
outflow by above considerations can be expected to have a mildly relativistic
bulk velocity, the observed spectrum will be affected by special relativistic
effects. In the next section, we will construct a light curve model taking these
effects into account.

\section{Outburst light curve model}\label{sc:model}

We can obtain the time evolution of the diffusive luminosity from the
time-dependent photon diffusion equation. The result in A80, used in
I98, was derived using this approach. The derivation assumes a
radiating spherical cloud undergoing linear homologous expansion, with
$R(t) = R_0 + t \dot{R}_0$,
where $R$ represents the radius of a photosphere, from
which the photons are observed to emanate.
In the following, we briefly summarize the derivation of the bolometric
luminosity presented in A80 and \citet{arnett1982}, without the assumption of
linear expansion from the outset.
The result is then complemented by taking into account optically thin
radiation, and special relativistic corrections to geometry, time delays and
optical thickness. Finally, we present plots demonstrating the time evolution
of the luminosity and spectra, and derive analytic approximations for maximum
luminosity, flux and spectral peak. The main difference to the LV96 model is the
contribution to the luminosity from the photon diffusion, assumption of
linear expansion and the explicit calculation of the time-dependent flux.

\subsection{Bolometric luminosity}\label{sc:luminosity}

From the first law of thermodynamics and the diffusion approximation, a
separable partial differential equation that describes the thermal
evolution of a homologously expanding sphere is obtained (A80)
\begin{equation}
    4T^4\left(\frac{\dot{T}}{T} + \frac{\dot{V}}{V}\right) = \frac{1}{r^2} \frac{\partial}{\partial r}
    \left( \frac{c}{3\kappa \rho} r^2 \frac{\partial T^4}{\partial r} \right).
\end{equation}
This can be solved by separation of variables. Set $x=r/R(t)$, and
substitute
\begin{equation}\label{eq:Tsubstitution}
    T(x, t)^4 = \psi(x) \phi(t) T_0^4 R_0^4/R(t)^4
    = \psi(x) \phi(t) T_0^4 \expfac(t)^{-4},
\end{equation}
where $T_0=T(0,0)$, so that the adiabatic component of temperature evolution, $T \propto
R(t)^{-1}$, is factored out.
Similarly, density can be written as
$\rho(r,t)=\rho_0\eta(x)\expfac(t)^{-3}$, where $\eta(x)$ parametrizes the
density profile. In general, the opacity $\kappa$ depends on position,
temperature, and composition. To obtain separability, we are
restricted to $\kappa = \kappa(x)$, which can be subsumed into
$\eta(x)$.
This is equivalent to assuming that
the primary contribution to opacity is from Thomson scattering, which
for the initial diffusion dominated part of the outburst is a valid
assumption.

Subsituting $T$ and $\dot{V}/V = 3\dot{R}/R$
we obtain a separated form
\begin{equation}\label{eq:diffODE}
    \frac{\alpha\tdiff R_0}{R(t)} \frac{\dot{\phi}(t)}{\phi(t)} =
    \frac{1}{x^2\psi(x)} \frac{\partial}{\partial
    x}\left(\frac{x^2}{\eta(x)}\frac{\partial\psi}{\partial x}\right)
    = -\alpha,
\end{equation}
where $\tdiff = 3 R_0^2\kappa_T\rho_0/(c\alpha)$ is the diffusion time-scale and
$\alpha$ is a dimensionless constant. This is an
eigenvalue problem for both the spatial part $\psi$ and the temporal
part $\phi$, with $\alpha$ as the eigenvalue.
The spatial part, $\psi(x)$, can be explicitly solved when $\eta(x)$ is
constant and boundary conditions are chosen suitably.
The boundary conditions are $\partial_x\psi(0)=0$ in the centre, and
$\psi(a)=-(2/3)\partial_x\psi(x)/(\kappa\rho(x)R(t))|_{x=a}$ from the Eddington
approximation at the outer boundary $x=a$, corresponding to $\tau = 0$.
For the optically thick case, this reduces to $\psi(1)=0$.
In this case, the solution for the spatial part is
\begin{equation}\label{eq:psisol}
    \psi(x)
    = \frac{\sin\left(x\sqrt{\alpha}\right)}{x\sqrt{\alpha}},
\end{equation}
where the normalization $\psi(1) = 0$ has been used.

A solution for $\phi(t)$, with an arbitrary radial evolution $R(t)$,
is
\begin{equation}\label{eq:phisol}
    \begin{split}
        \phi(t) &=
        \exp\left( -\frac{c\alpha}{3 R_0^3\kappa_T\rho_0} \int_0^t R( t') \ud t' \right) \\
        &= \exp\left( -\frac{1}{\tdiff R_0} \int_0^t R( t') \ud t' \right).
    \end{split}
\end{equation}

The total mass within $x$ is given by
\begin{equation}\label{eq:totmass}
    M(x) = \int_0^{xR}\rho 4\pi r^2\ud r 
    = 4\pi\rho_0 R_0^3I_M(x),
\end{equation}
where $I_M(x) = \int_0^x \eta(x') x'^2\ud x'$ is a dimensionless factor,
which depends on the density distribution. For a constant density distribution 
$I_M(x)=x^3/3$.
Similarly, the total thermal energy content within $x$ for a radiation dominated case is
\begin{equation}
    E_T(x,t) = \int_0^{xR} 4\pi r^2 aT^4\ud r
    = 4\pi R_0^3 a T_0^4 \phi(t) \expfac(t)^{-1} I_T(x),
\end{equation}
where $I_T(x) = \int_0^x \psi(x') x'^2\ud x'$ is likewise dimensionless,
and depends on the temperature distribution.
Integrating the spatial part of equation \eqref{eq:diffODE} from 0 to $x$ gives
\begin{equation}
    \psi'(x) = -\alpha\eta(x)I_T(x).
\end{equation}
Using these results, the luminosity directed outwards at a distance $x$ from the
centre can be written in several equivalent forms,
\begin{equation}\label{eq:Ldiff}
    \begin{split}
        \Ldiff(x,t)
        &= 4\pi r^2\frac{cl}{3}\left(-\frac{\partial aT^4}{\partial r}\right) 
        = x^2\frac{E_T(x,t)}{\tdiff}\expfac(t) \\
        &= x^2\frac{E_T(x,0)}{\tdiff}\phi(t) \\
        &= \frac{4\pi R_0c}{3\kappa}\frac{x^2E_T(x,0)}{M(x)}\alpha I_M(x)\phi(t),
    \end{split}
\end{equation}
where $l = [\rho(x)\kappa]^{-1}$ is the photon mean free path.%
\footnote{
    For $x=1$, this
    result reduces to the form used in I98 if
    we assume $\alpha=\pi^2$ and $R_0 = \rbh$, or the Bondi-Hoyle radius
    of the secondary black hole.
}
The surface luminosity is then just
\begin{equation}\label{eq:surflum}
    L(1,t) = \frac{E_T(1,0)}{\tdiff}\phi(t) =
    \frac{4\pi}{3}\frac{R_0c E_T(1,0)}{\kappa M}\alpha I_M \phi(t).
\end{equation}
Intuitively, the result shows that the luminosity is given by the
initial diffusive luminosity $E_T/\tdiff$ modulated both by radiative
and adiabatic cooling, given by $\phi(t)$.
A80 showed that the luminosity is only weakly dependent on the density profile
through $\alpha I_M$, and the initial diffusive luminosity is not affected much
by strong density gradients in the outflow. However, a steep gradient will start
becoming optically thin earlier. This will allow radiation to escape, and will
increase the luminosity over the diffusive limit.

Before discussing the optically thin part of the luminosity, we have to assess
some shortcomings in the model. The central problem, as noted in A80, is that
equation~\eqref{eq:psisol}, while producing a self-consistent model, is
not entirely accurate.
The actual temperature distribution after a strong radiative shock is
more homogenous \citep[see e.g.][]{elliott1960}. 
This places more of the thermal energy near the surface
and produces an initial transient increase in the luminosity. During later
evolution the luminosity is
suppressed by a constant factor of $\bigO(1)$, as presented in A80. We
wish to provide a conservative estimate for the luminosity, so we will not
include the transient.
However, the unphysical temperature profile will affect the optical depth in the
outer parts of the outflow envelope and also the spectral shape. To mitigate
this, in the following we will assume $\psi(x)=1$, which is more
physical, but also not strictly self-consistent.
We further assume $\eta(x)=1$, which also implies $\alpha=\pi^2$. This is not an
unduly bad approximation for the initial shocked volume.  Furthermore, the
initial density gradient does not
affect the diffusive luminosity much, as shown in A80.  At later times,
homologous expansion will cause a gradient $\eta(x)\propto (x\expfac)^{-3}$ to
form, which would increase the optically thin luminosity, but in our case, most of
the energy is radiated away during the early outburst. See \citet{falk1977},
A80 and \citet{arnett1982} for more discussion about the behaviour and effects
of the temperature and density profiles during the expansion.

We can estimate the optically thin contribution by assuming that all
radiation from regions with $\max(\taus,\taueff)\leq\tau_p$ escapes. Here $\tau_p$ is
a limiting photospherical optical depth. We take $\tau_p=2/3$, even
though this value is strictly valid only for plane parallel geometries.
For pure scattering, this optical depth corresponds to a relative
distance $x_\sigma$ implicitly defined by
\begin{equation}
    \tau_p = \int_r^R \rho(r)\kappa_T\ud r =
    \expfac(t)^{-2}\tau_{\sigma,0} \int_{x_\sigma}^1\eta(x')\ud x'.
\end{equation}
With the assumptions made so far,
\begin{equation}\label{eq:xs}
    x_\sigma(t) = 1-\frac{\tau_p}{\tau_{\sigma,0}}\expfac(t)^2.
\end{equation}
During later parts of the evolution, we may have $\taueff>\taus$, and as
such $x_\sigma < x_e$, where $x_e$, defined in the next section, is the
relative radial position of the photosphere as determined by effective optical depth.
As such, we take the extent of the photosphere to be
\begin{equation}\label{eq:xp}
    x_p(t) = \max(x_\sigma(t), x_e(t)).
\end{equation}

The bolometric luminosity $\Lthin$ of the escaping radiation can then be
estimated as
\begin{equation}\label{eq:Lthin}
    \begin{split}
    \Lthin(t) &= 4\pi R_0^3 aT_0^4 \phi(t)\expfac^{-1}x_p^2\psi(x_p)\left(-\dot{x}_p\right) \\
        &= 4\pi R_0^3 aT_0^4 \phi(t)x_p^2\psi(x_p)\frac{2\tau_p}{\tau_{\sigma,0}}\dot{\expfac}
    \end{split}
\end{equation}
where the second equality is valid for the case $x_\sigma > x_e$. When the outflow
cools, the effective optical depth will at some point start to increase, and a
situation where $\dot{x}_p > 0$ may result. From this point onwards, we take the
optically thin luminosity to be zero.
Compared to a purely diffusive case, the escaping radiation will result in
$\psi(x)$ decreasing smoothly
to $\sim$zero when $x>x_p$. We will use an approximation with $\psi(x) = 0$
for $x\geq x_p$. As such, the diffusive luminosity must be
calculated at $x=x_p(t)$ during the early part of the light curve, and at
$x=\min_t x_p(t)$ during the late part, if $\dot{x}_p > 0$.
Otherwise an unphysical increase in the luminosity would result when the cooling
outflow gets optically thicker.
With these considerations, we
arrive at the estimate for total luminosity, 
\begin{equation}\label{eq:Ltot}
    \Ltot(t) = \Ldiff(x,t)+\Lthin(t).
\end{equation}

Our final consideration is the rising part of the light curve for $t<0$. The
radiative energy $E_T$ in the outflow is generated by the cooling matter, with
some cooling time-scale $t_c$. \rv{As an upper bound, the
bremsstrahlung cooling time-scale from equation~\eqref{eq:fftime-scale}
can be used.}
This is much less than the the impact time-scale $\tdyn$ or the two defining
time-scales of the luminosity evolution, $\tdiff$ and $\thyd$. As such, the
initial radiative energy is produced at a nearly constant rate during the impact time $\tdyn$,
and we could in principle set $E(x,t) = (1+t/\thyd)E(x,0)$ for $t < 0$.
However, since the maximum luminosity and flux are obtained at $t\geq 0$ and the 
A80 model is not constructed to model evolution before $t=0$, we set
$E(x,t) = E(x,0)$ for $t<0$.

\subsection{Emitted spectrum}

We are mostly interested in the continuum radiation and its time evolution.
This is natural, since a successful modelling of the line radiation is only
feasible numerically.
For both the diffusive, optically thick case and the optically thin case, the spectrum of
the radiation is in given by a blackbody with an effective
temperature corresponding to the radial depth given by $\taueff=2/3$, or the
surface of last emission. After the last emission,
the photons are scattered by the electrons several times, but as noted in
Section~\ref{sc:lvredux}, the spectrum is not appreciably comptonized by this.
For a careful analysis, as noted in \citet{shibata2014}, the effective
optical depth $\taueff$ needs to be modified
for even mildly relativistic bulk velocities. The principal reason is
that while the scattering events are isotropic in the flow rest frame,
they are nonisotropic in the observer frame. In the optically thick
scattering dominated case this is observable even for mildly
relativistic flows, since the small effect is multiplied by the large
number of scatterings. For completeness, we will first discuss a
solution for a general $\psi(x)$.
\citet{shibata2014}
obtain the following effective optical depth
\begin{equation}\label{eq:shibata}
    \taueff =
    \left[\frac{2}{3}\Gamma^2(\beta^2+3)+(\Gamma\beta)^2\frac{\taus}{\taua}\right]^{-1/2}
    \frac{\sqrt{\taua(\taua+\taus)}}{\Gamma(1-\beta\cos\theta)},
\end{equation}
where $\Gamma = (1-\beta^2)^{-1/2}$, $\beta=v/c$, and $\theta$ is the angle
between the bulk flow velocity $v$ and the line of observation. The
optical depths $\taus(x)$ and $\taua(x)$ are given by
\begin{gather}
    \taus(x) = \tau_{\sigma,0}\expfac^{-2}\int_x^1\eta(x')\ud x' \\
    \taua(x) = \tau_{a,0}\expfac^{-3/2}\phi^{-7/8}\int_x^1 \eta(x')\psi(x')^{-7/8}\ud x',
\end{gather}
where $\tau_{\sigma,0} = R_0\rho_0\kappa_T$ and
$\tau_{a,0} = R_0K_a\rho_0^2T_0^{-7/2}$, from
equation \eqref{eq:kramer}.
For the thermal energy profile \eqref{eq:psisol}, the integral
$\int_x^1\psi(x')^{-7/8}\ud x'$
could be approximated by  $8(1-x)^{1/8}$ which is accurate to
$\sim4$ per cent within $x\in[0,1]$.
However, as discussed above, a steep thermal profile is not very
physical, and the small temperatures in the region near $x=1$ lead to
unphysically high opacities, so the approximation $\psi(x)=1$ is
used instead.
We can now solve $x_e$ from $\taueff(x_e)=\tau_p$. For arbitrary
density and temperature profiles this has to be done numerically, but for
constant temperature and density profiles we find
\begin{equation}\label{eq:xa}
    x_e(t) = 1 - \frac{\tau_p}{\taueff(t)},
\end{equation}
where $\taueff(t)$ is calculated from equation \eqref{eq:shibata} by evaluating
$\taus(x)$ and $\taua(x)$ at $x=0$.

In our assumption we are observing an
optically thick photosphere with minimal comptonization.
The spectrum for total luminosity
is given by a blackbody with an effective temperature $\Teff$
\begin{equation}\label{eq:Teff}
    \Teff = \left(\frac{\Ltot(t)}{4\pi[x_p(t)R(t)]^2\sigma_B}\right)^{1/4},
\end{equation}
where the evolution is modified from a simple optically thick model by
the relative strength of the scattering and absorption
processes through $x_p$.

\subsection{Relativistic effects}

For a spherical photosphere expanding at a sufficiently high velocity,
the observed geometry and spectrum will be affected by special relativistic corrections.
We will briefly introduce the most significant
of these effects, which we include in the model. Discussions of
special relativistic effects in expanding fireballs can be found in
e.g. \citet{ryde2002} and \citet{peer2008}.
An important simplifying assumption
we make is to take the expansion law $R(t)$ to be exactly linear,
\begin{equation}\label{eq:modelR}
    R(t) = R_0\left(1 + \frac{c\beta}{R_0}t\right)
    = R_0\left(1 + \frac{t}{\thyd}\right),
\end{equation}
where the hydrodynamical time-scale is now $\thyd = R_0/(c\beta)$.
We also assume that the expansion velocity $x_p \beta c$ of the photosphere is
nearly constant.
This is reasonable in light of the discussion in
Section~\ref{sc:discussion1}, and the fact that the relative radius of the
photosphere, $x_p$, does not vary rapidly. 

\begin{figure}
    \includegraphics[width=\columnwidth]{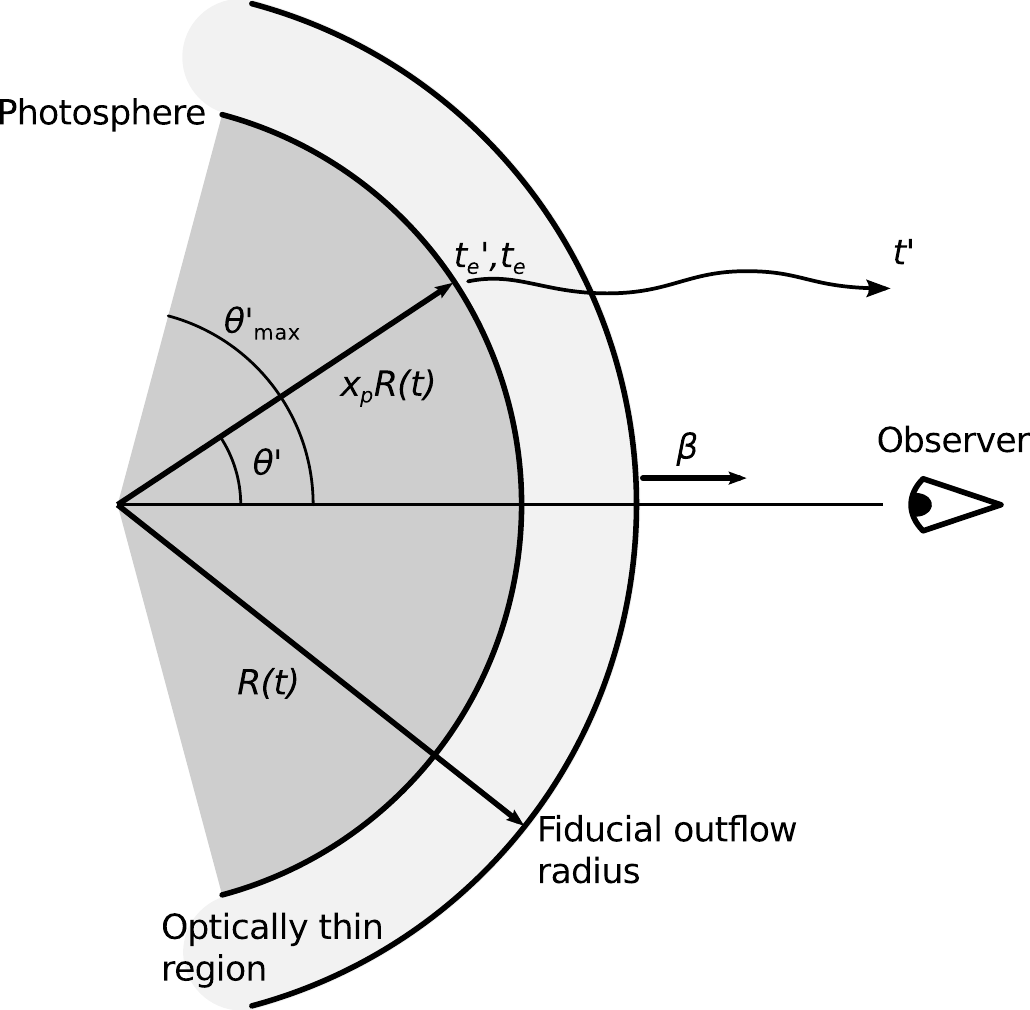}
    \caption{\label{fig:outflow_geometry}
        A sketch of the outflow geometry within $\theta'_\text{max}$. The
        outflow radius $R(t)$ and the radius of the photosphere $x_p R(t)$ are
        shown. Also shown are the times of emission in the outflow rest frame
        $t_e$ and the observer's frame $t'_e$, and the
        time of observation $t'$ in the observer's frame. 
        The light grey shading indicates the
        optically thin region of the outflow, and the dark grey shading
        indicates the optically thick region.
    }
\end{figure}

The radiation from elements of the photosphere moving from the projected centre
with different angles $\theta'$ with respect to the observer at $\theta'=0$ will
be observed at different times due to
geometrical and special relativistic effects. We define the zero point of the time of observation
$t'$ so that when $t'=0$, the emission time $t_e$ in the frame of the
photosphere element is $t_e=0$, for $R = R_0$ and $\theta'=0$. With this
definition,
\begin{equation}
    t_e(t', \theta') = \frac{D(\theta')}{1+z}\left[t'-\frac{R_0}{c}(1-\cos\theta')\right],
\end{equation}
where $D(\theta) = \Gamma^{-1}(1-\beta\cos\theta)^{-1}$ is the Doppler
boosting factor, and primes denote quantities defined in the rest frame of the observer.
The second term in square brackets is the geometrical correction for light
travel time from different latitudes.
In addition, instead of a hemisphere, the observer will only see a conical
segment of the photosphere, given by the maximum latitude
$\theta'_\text{max}$,
\begin{equation}\label{eq:thetamax}
    \theta'_\text{max} \leq \arccos(\beta)
\end{equation}
where $\theta' = 0$ corresponds to the direction towards the observer. The
angles and the outflow geometry are illustrated in
Figure~\ref{fig:outflow_geometry}.

Equation \eqref{eq:thetamax} is only approximate in the sense that it
depends on $x_p(t)$ changing slowly with time, which is a good
approximation in our case.
The observed spectral flux of the element is
\begin{equation}
    I'_{\nu'}(t',\theta') = D(\theta')^3 I_{\nu}(t(t',\theta')),
    \quad \nu = \frac{(1+z)\nu'}{D(\theta')},
\end{equation}
where $z$ is the source redshift. The observed spectral
irradiance is obtained by integrating over the visible section of the
photosphere, and we find
\begin{equation}\label{eq:flux}
    F'_{\nu'}(t') = \frac{\pi(1+z)}{D_L^2}\int_{0}^{\theta'_\text{max}}
    I'_{\nu'}(t',\theta)[x_p(t_e)R(t_e)]^2\sin(2\theta)\ud\theta.
\end{equation}
Equation \eqref{eq:flux} connects the model to observations.

\subsection{Final model and analytic approximations}\label{sc:final_model}

The model is now completely determined
and specified by the following parameters: initial radius $R_0$,
temperature $T_0$, density $\rho_0$ and expansion velocity
$\beta$. Of these, $R_0$,
$T_0$ and $\rho_0$ can be estimated with the results in
Section~\ref{sc:disc_impact}. These parameters set
the maximum diffusive luminosity.
With $\rho_0$, the parameter $\beta$ sets the time-scale of the
outburst and also affects the optically thin luminosity.
Both are difficult to estimate from first principles, but for
the expansion velocity a conservative estimate
$\beta \sim \vrel/c$ can be used. The total bolometric luminosity is given
by equation \eqref{eq:Ltot}, with the diffusive and optically thin
components determined from \eqref{eq:Ldiff} and \eqref{eq:Lthin}. The
spectrum is derived from \eqref{eq:flux}, with the rest frame $I_\nu$
given by a blackbody spectrum with effective temperature specified by
equation \eqref{eq:Teff}. We give a link to the computer code for performing
the model computations in Appendix~\ref{sc:code}.

The full model is somewhat complex, but we may make some crude
estimates for the luminosity, optical flux
and the spectral peak.
The asymptotic behaviour of the bolometric luminosity is mainly
determined by the function $\phi(t)$, which for an expansion law
of the form of equation \eqref{eq:modelR} is
\begin{equation}
    \begin{split}
        \phi(t) &=
    \exp\left\{-\frac{1}{2}\frac{\thyd}{\tdiff}
    \left[\left(1+\frac{t}{\thyd}\right)^2 - 1\right]\right\} \\
    &= \exp\left[-\frac{1}{2}\frac{\thyd}{\tdiff}
    \left(\expfac(t)^2 - 1\right)\right].
    \end{split}
\end{equation}
For $t\sim 0$, $R\sim R_0$ we have
\begin{equation}
    \phi(t) = 1-\frac{t}{\tdiff}+\bigO(t^2),
\end{equation}
and for $t\fromto\infty$, we have
\begin{equation}
    \phi(t) \fromto \exp\left[-\frac{1}{2}\left(\frac{t}{\sqrt{\tdiff\thyd}}\right)^2\right].
\end{equation}
As a rough estimate, we can say that the luminosity will show an initial
linear decline determined purely by diffusion through $\tdiff$. At later
times, we would observe an exponential decrease characterised by a time-scale
$\sqrt{\tdiff\thyd}$. Relativistic effects increase the maximum
observed luminosity and decrease the observed outburst duration by $1+\beta$ to first order.
The intrinsic maximum luminosity, obtained at $t=0$,
and given by equation \eqref{eq:surflum}, is
\begin{equation}\label{eq:Lmaxestimate}
    \begin{split}
        \Lmax &= \frac{E_T(1,0)}{\tdiff} \\
              &= 7.2\times 10^{45} \pms^{2/3} \ph^{1/3} \pr^{-1/3}\incp^{2/3} \,\si{erg.s^{-1}},
    \end{split}
\end{equation}
using the estimates and the normalization of Section~\ref{sc:disc_impact}.

If we assume optical thickness (i.e.\ pure diffusion, $x_p(t)=1$),
$\beta\sim 0$, so that $R(t)$
and $\Teff(t)$ are nearly constant with respect to $\theta'$, and that $R_0/c\sim 0$, 
we can explicitly solve for the observed flux in the
Rayleigh--Jeans limit.
Observed Rayleigh--Jeans flux is given by
\begin{equation}\label{eq:RJflux}
\begin{split}
    F'_{\nu'}(t') &=
    \frac{(1+z)^3 \pi}{D_L^2} \frac{2 \nu'^2}{c^2} k\Teff(0) R_0^2 \\
&\quad\times \int_{0}^{\theta'_\text{max}} D(\theta)\expfac(t_e)^{3/2}\phi(t_e)^{1/4}\sin(2\theta)\ud\theta \\
&\sim F'_{\nu'}(0)\left(1+\frac{2\beta}{3}\right)\expfac(t_e)^{3/2}\phi(t_e)^{1/4},
\end{split}
\end{equation}
where $t_e = t'(1+\beta)/(1+z)$. The result is an increase in flux by a factor
of $1+2\beta/3$ from the non-relativistic value, but with a corresponding
decrease in outburst duration by a factor of $1+\beta$. The maximum flux is
determined by the maximum of $\xi^{3/2}\phi^{1/4}$,
\begin{gather}
    [\xi^{3/2}\phi^{1/4}](t_\text{max}) =
        \left(\frac{6\tdiff}{\thyd}\right)^{3/4}\exp\left(-\frac{3}{4}
            + \frac{\thyd}{8\tdiff}\right), \\
    t_\text{max} = \sqrt{6\thyd\tdiff} - \thyd.
\end{gather}
Assuming $\thyd\ll\tdiff$, we find
\begin{equation}\label{eq:RJmaxflux}
\begin{split}
    \frac{F'_{\nu',\text{max,RJ}}}{(1+z)^3}
    &\sim 30 \beta^{3/4}\left(1+\frac{2\beta}{3}\right)\,\si{mJy} \\
    &\quad\times\ph^{5/6} \pms^{5/3} \pn^{3/4} \pr^{17/12}\incp^{-17/6}
    \plumdist^{-2} \nu_{550},
\end{split}
\end{equation}
where $\nu_{550} = \nu'(c/550\,\si{nm})^{-1}$.
The time it takes to reach the maximum flux is
\begin{equation}\label{eq:RJmaxfluxtime}
    t'_\text{max} \sim 2.97\times 10^{6}\ph^{1/2}\pms\pn^{1/2}\pr\incp^{-2}
    \frac{1+z}{\sqrt{\beta}(1+\beta)}\,\si{s}.
\end{equation}

\begin{figure*}
    \includegraphics[width=\textwidth]{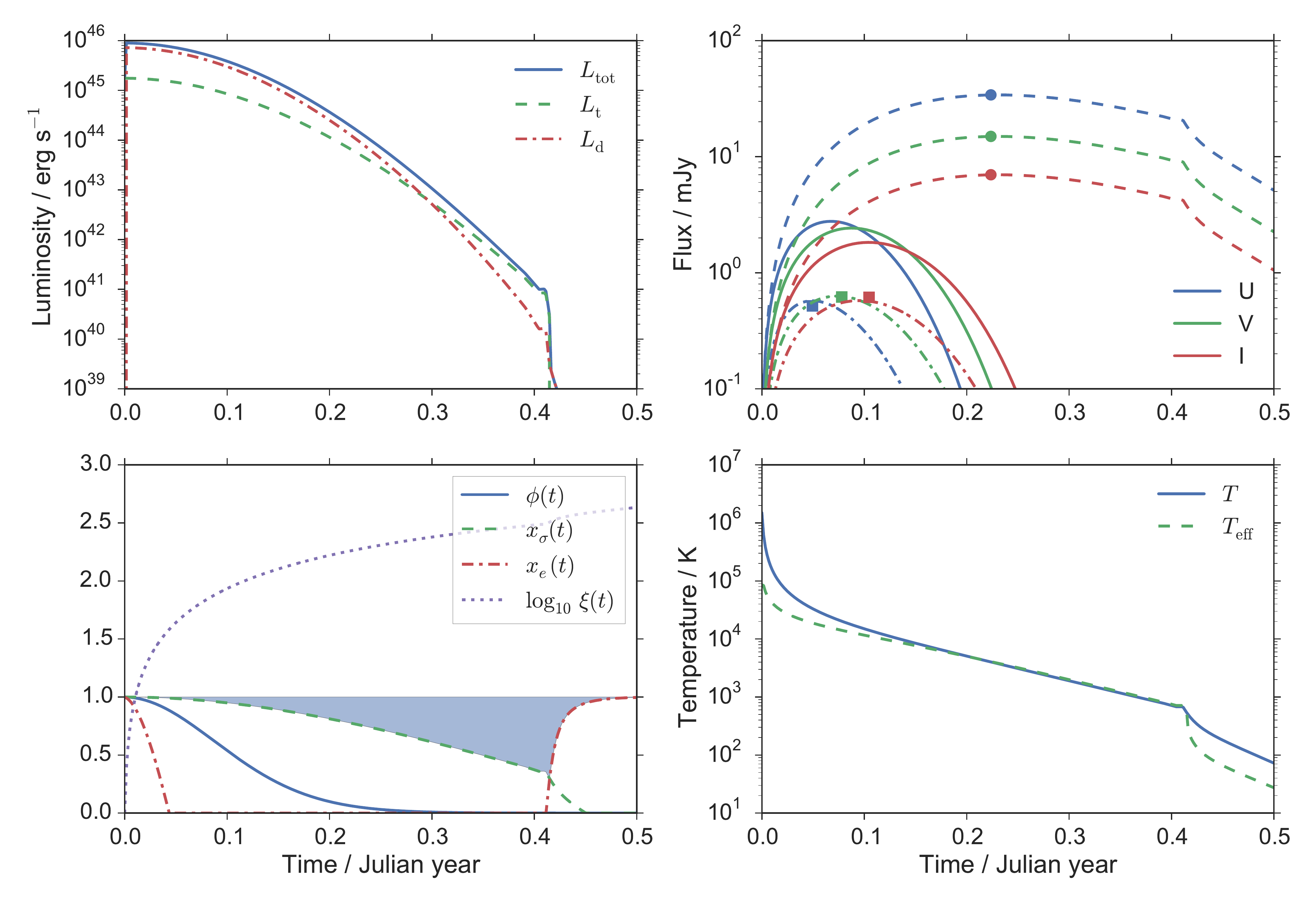}
    \caption{\label{fig:modelplot}
        Simulated quantities for an outburst with nominal parameters (see
        Section~\ref{sc:disc_impact}) and $\beta = 0.2$. 
        Upper left panel
        displays total, diffusive and optically thin luminosities ($\Ltot$, $\Ldiff$ and
        $\Lthin$, respectively).
        Upper right panel shows optical fluxes in $U$,
        $V$ and $I$-bands (assuming the nominal luminosity distance of
        $D_L=1500\,\si{Mpc}$). Model output is indicated with solid lines, the
        estimates from equations \eqref{eq:RJflux}, \eqref{eq:RJmaxflux} and
        \eqref{eq:RJmaxfluxtime} with dashed lines and filled circles, and
        estimates from equations \eqref{eq:BBflux} and \eqref{eq:BBmaxflux} with
        dash-dotted lines and filled squares.
        Bottom left panel shows the time evolution of $\expfac$, $\phi$ and the
        relative photospheric radii $x_\sigma$ and $x_e$ corresponding to the
        electron scattering and effective optical thicknesses respectively. The
        shaded blue region indicates the optically thin extent at each time.
        Bottom right panel depicts time evolution of the model temperature $T$
        as defined by equation \eqref{eq:Tsubstitution} and the effective
        temperature $\Teff$ defined by \eqref{eq:Teff}.
    }
\end{figure*}

Using the same assumptions, we can also make estimates for the spectral peak
and the flux at the peak for a black body. The position of the spectral
peak is proportional to the effective temperature, so
\begin{equation}\label{eq:BBpeak}
\begin{split}
    \nu'_\text{peak,BB} 
    &\sim 5.879\times 10^{10}\, \frac{1+\beta}{1+z}\Teff(0)\, \expfac(t_e)^{-1/2}\phi(t_e)^{1/4}\,\si{Hz} \\
    &= 7.2\times 10^{15}\,\frac{1+\beta}{1+z} \,\si{Hz} \\
       &\quad\times\ph^{-1/12}\pms^{-1/6}\pr^{-5/12}\incp^{5/6} \expfac(t_e)^{-1/2}\phi(t_e)^{1/4}.
\end{split}
\end{equation}
Based on this we can say that
\eqref{eq:RJmaxflux} and \eqref{eq:RJmaxfluxtime} are usable only in the limit $\nu' \ll
\nu'_\text{peak,BB}$.
The flux at peak is
\begin{equation}\label{eq:BBpeakflux}
\begin{split}
\frac{F'_{\nu'_\text{peak},BB}}{1+z} &\sim
    \frac{\pi}{\lumdist^2}\,1.896\times
    10^{10}\,\si{mJy.K^{-3}}\,\Teff(0)^3 R_0^2 \\
        &\quad\times \expfac(t_e)^{1/2}\phi(t_e)^{3/4}
    \int_0^{\theta'_\text{max}}D(\theta)^3\sin(2\theta)\ud\theta \\
    &\sim 0.23\,(1+2\beta)\,\si{mJy}\\
    &\quad\times \ph^{5/12}\pms^{5/6}\pr^{1/12}\incp^{-1/6}\plumdist^{-2} \\
        &\quad \times\expfac(t_e)^{1/2}\phi(t_e)^{3/4}.
\end{split}
\end{equation}
In principle, we can also estimate the maximum flux using the black body assumption,
if we additionally assume $\thyd \ll \tdiff$ and $\beta\sim 0$. In this case,
the flux is given by
\begin{equation}\label{eq:BBflux}
    F'_{\nu'}(t') = \frac{(1+z)\pi }{D_L^2} R(t_e)^2 B_{(1+z)\nu'}(\Teff(t_e)).
\end{equation}
The maximum flux is
\begin{equation}\label{eq:BBmaxflux}
    \begin{split}
\frac{F'_{\nu',\text{max,BB}}}{(1+z)^4}
&= \frac{\pi}{\lumdist^2} 
\frac{R(t_e)^2 2h\nu'^3/c^2}{\exp\left[\frac{(1+z)h\nu'}{k\Teff(0)}\expfac(t_e)^{1/2}\phi(t_e)^{-1/4}\right]-1}
\\
&\sim 1.6\times 10^{-3}\,\si{mJy}\,f(\expfac)\\
        &\quad
\times\pms^{4/3}\ph^{2/3}\pn^{1/2}\pr^{4/3}\incp^{-8/3}\plumdist^{-2}\nu_{550}^3,
    \end{split}
\end{equation}
where 
\begin{align}
    f(\expfac) &= \expfac^2 \left\{ \exp\left[ 
C_1 \xi^{1/2} \exp(-C_2 (\expfac^2-1))^{-1/4}\right] - 1\right\}^{-1}\\
C_1 &= 0.21(1+z)\pms^{1/6}\ph^{1/12}\pr^{5/12}\incp^{-5/6}\nu_{550} \\
C_2 &= 1.7\times 10^{-5} \beta^{-1}
    \pms^{-2/3}\ph^{-1/3}\pn^{-1}\pr^{-2/3}\incp^{4/3}.
\end{align}
This flux maximum is determined by maximum of $f$, for which no convenient
analytic result exists. For nominal parameter values \rv{and $\beta=0.2$} we can numerically find 
$\expfac_\text{max}\sim \mathrv{72}$, which gives $t_\text{max}\sim
\mathrv{72}\thyd$ and $f(\expfac_\text{max})\sim \mathrv{430}$.
Thus, the estimate for nominal $V$-band maximum brightness is $\sim
\mathrv{1.8}\,\si{mJy}$.

\begin{figure}
    \includegraphics[width=\columnwidth]{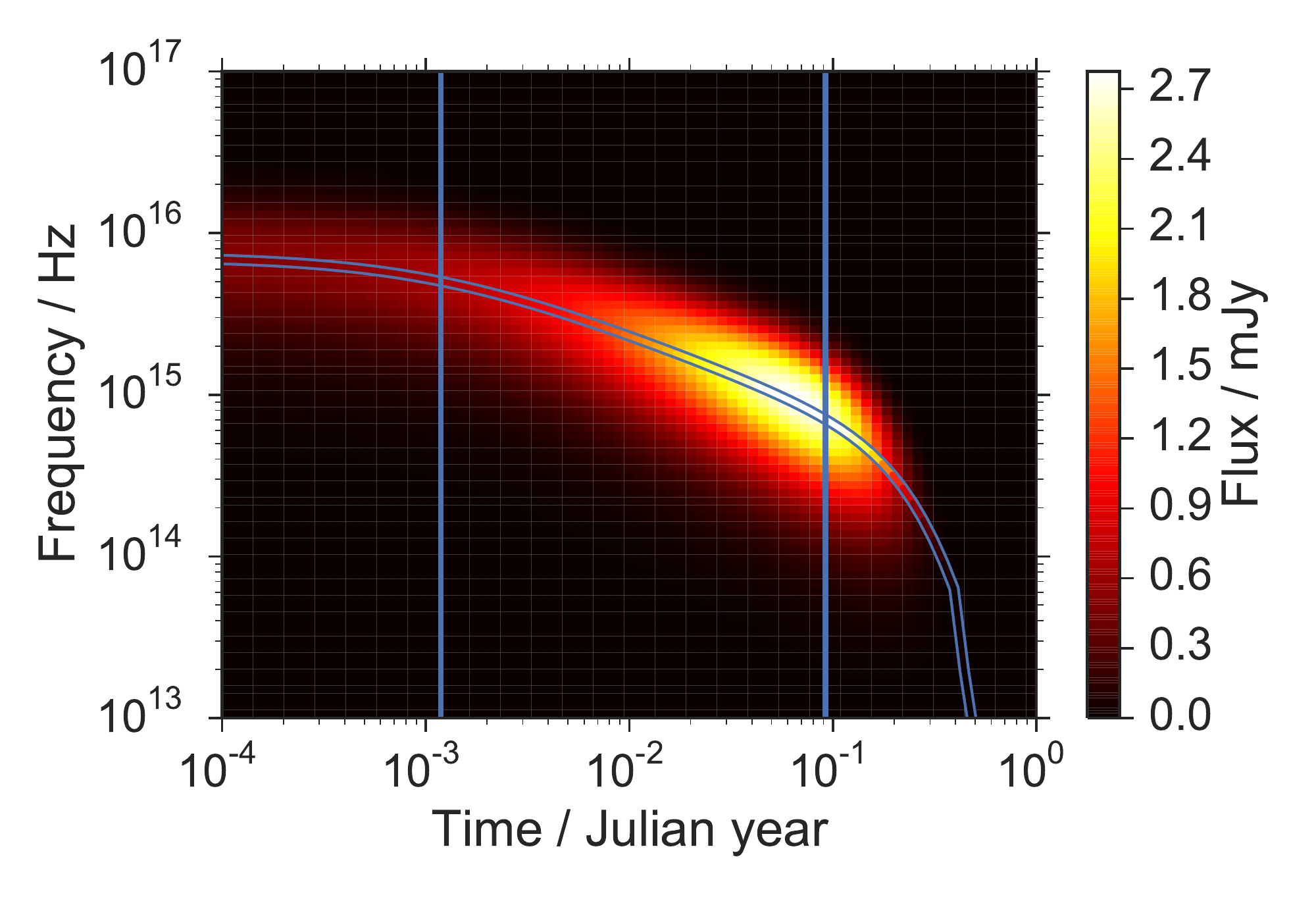}
    \caption{\label{fig:spectrumplot}
        Time evolution of the observed spectrum for
        an outburst with nominal parameters and $\beta = 0.2$. Note the
        logarithmic scale on the $x$-axis.
        The curved blue lines bracket the estimate of the spectral peak position, given by
        equation \eqref{eq:BBpeak}, and the flux at spectral peak, given by
        equation \eqref{eq:BBpeakflux}. The two vertical lines indicate the
        time-scales $\thyd$ and $\sqrt{\thyd\tdiff}$ from left to right,
        respectively.
    }
\end{figure}

We can compare these estimates to Figures~\ref{fig:modelplot}
and~\ref{fig:spectrumplot}, which depict the model results for an example
outburst with nominal parameter values and a moderate value of $\beta=0.2$.
The effect of Doppler boosting through $\beta$ can be clearly seen in
the flux light curves when compared to the blackbody estimate of equation
\eqref{eq:BBflux}. Otherwise, using the naive blackbody value for flux is a
good approximation.
On the contrary, the Rayleigh--Jeans estimates from
\eqref{eq:RJmaxflux} and \eqref{eq:RJmaxfluxtime} are inaccurate at optical
wavelenghts. The
estimate for maximum flux is too high by a factor of few.
Likewise, the estimated time-scale of the outburst is too long. This is not
surprising, since during the brightest part of the outburst, the spectral peak
is at optical frequencies as well, and the Rayleigh--Jeans assumption does not
hold.

The Figure~\ref{fig:spectrumplot} clearly visualizes how
the initial spectral evolution is dominated
by the hydrodynamical time-scale through expansion and adiabatic
cooling, resulting in an increasing peak flux at smaller peak
frequencies.
After $t\sim\sqrt{\tdiff\thyd}$, photon diffusion cools the gas and the maximum
flux and the peak frequency diminish quickly. The estimates given by equations
\eqref{eq:BBpeak} and \eqref{eq:BBpeakflux} are reasonably good.

During most of the outburst, the diffusive luminosity dominates the optically
thin luminosity by approximately an order of magnitude. At later times,
when the bubble has rarefied enough, the optically thin luminosity exceeds the
diffusive contribution. However, soon after this, the low temperature induces a
quick rise in the effective optical depth, quenching the optically thin
contribution to luminosity. It is clear that for nominal parameter values, the
optically thin luminosity is only a minor component of the total luminosity.
However, as noted in Section~\ref{sc:luminosity}, a pre-existing or expansion
induced density gradient could enhance the optically thin contribution.

\section{Accretion disc impacts as an observational tool}%
\label{sc:observational_tool}

When the outburst model is tied to
physical parameters at the impact site through estimates in
Section~\ref{sc:disc_impact}, we can use the model to probe the physical
parameters of the accretion disc through $\ph$ and $\pn$, and the
orbital parameters and mass of the impactor through $\pr$, $\pms$ and
the outburst timings. In constructing the model, the nominal values of these
parameters were chosen to represent the scenario originally presented in LV96.
Since the mass of the primary in LV96 model and its later developments is high,
$\sim 10^{10}\usk\Msun$, nominal parameter values are not representative of the
general SMBH population.

To properly estimate the possibility of observing accretion disc impact
outbursts, we construct a large number of impact scenarios based on
observational data of quasars. Quasars form a convenient benchmark, since it is
well established that they represent a population of SMBHs, and that their
luminosity is powered by matter accretion through a disc \citep[e.g.][]{malkan1983}.
Furthermore, quasar activity is commonly linked to galaxy merger events --
e.g. \citet{johansson2009,treister2012}, but see also \citet{cisternas2011} -- which makes the
existence of a close binary of SMBHs within a quasar a plausible scenario.
Indeed, there are now several promising candidates detected with a variety of
methods \citep{sil1988,graham2015a,liu2015,graham2015b}.
We compare the maximum outburst luminosity and flux calculated using our
accretion disc impact model with the quiescent (baseline)
luminosity and flux of the quasar to assess how likely it is
that the outburst could be observed.

For our data, we use the observations of 62185 SDSS quasars in \citet{steinhardt2010}.
This data yields the mean masses and luminosities in 9 evenly spaced
redshift bins from $z=0.2$ to $z=2.0$. The paper also presents evidence
for a redshift-dependent luminosity limit for quasars of the
form
\begin{equation}\label{eq:subedd}
    \log L_\text{max} = \alpha(z) \log M + \log L_0(z),
\end{equation}
which the authors dub the sub-Eddington limit,
due to the result that $\alpha < 1$ in all the observed redshift bins.
For each redshift bin, we construct
a sample of quasars by sampling the black hole masses from a normal
distribution assuming the given mean and standard deviation for each bin.
The quasar luminosities are then set to the maximum expected luminosity
given by the equation~\eqref{eq:subedd} to yield conservative estimates
when compared to the accretion disc impact luminosities. The quasar redshift is 
generated from a uniform distribution over the redshift bin size.
We have reproduced the required values in Table~\ref{tb:steinhardt},
where it should be noted that for the redshift bin $z=0.6$--$0.8$, where
two sets of values are available, we have used the values derived using the H$\beta$ line.

\begin{table}
    \centering
    \caption{Mean quasar mass $\mean{M}$, mass standard deviation
        $\sigma_M$ and the sub-Eddington limit coefficients $\alpha$ and
    $L_0$ for each redshift bin, from \citet{steinhardt2010}.}
    \label{tb:steinhardt}
\begin{tabularx}{\columnwidth}{lcXXX}
    \toprule
    Redshift & $\mean{\log (M/\Msun)}$ & $\sigma_M$ & $\alpha$ & $L_0$ \\
    \midrule
0.2-–0.4  &  8.27 &  0.44 & 0.37  & 42.63 \\
0.4-–0.6  &  8.44 &  0.42 & 0.45  & 42.13 \\
0.6-–0.8  &  8.69 &  0.39 & 0.60  & 41.07 \\
0.8-–1.0  &  8.76 &  0.31 & 0.67  & 40.60 \\
1.0-–1.2  &  8.89 &  0.29 & 0.67  & 40.71 \\
1.2-–1.4  &  8.96 &  0.29 & 0.73  & 40.24 \\
1.4--1.6  &  9.07 &  0.28 & 0.68  & 40.66 \\
1.6--1.8  &  9.18 &  0.29 & 0.50  & 42.35 \\
1.8-–2.0  &  9.29 &  0.30 & 0.42  & 43.20 \\
    \bottomrule
\end{tabularx}
\end{table}

For each constructed quasar, we assume a standard Shakura--Sunyaev accretion
disc \citep{shakura1973} with $\alpha=0.1$,
which is consistent with observations \citep{king2007}.
The disc mass accretion rate is set to $\dot{M} = (L/\Ledd)\Medd$.
We then compute the model outputs using impact distances
$\pr=1$ and $\pr=10$, while setting $\incp=1$. The outflow velocity $\beta$ is
set equal to the impact velocity, $\beta=\vrel/c=(10\pr)^{-1/2}$. The mass ratio $q$ of the impacting
black hole to the quasar black hole is set in turn to $0.1$, $0.3$ and $1.0$.
The resulting maximum outburst luminosity is found at $t=0$, as outlined in
Section~\ref{sc:model}.
The maximum flux at $510\,\si{nm}$ (in observer frame)
is found by numerical maximisation procedure. This is compared against
the quasar flux at the same observed wavelength.
This is derived by using observed bolometric corrections in reverse.
We compute the observed flux from
\begin{equation}
F_{\nu} = \frac{1+z}{4\pi D_L^2}\frac{1}{\nu}\left(\lambda L_\lambda\right),
\end{equation}
where $\lambda = 510\,\si{nm}$ and $\nu = c/\lambda$. The value of
$\lambda L_\lambda$ is obtained using
the bolometric correction relations in \citet{runnoe2012}, which give 
\begin{equation}
\log(L_\text{iso}) = 4.89 + 0.91\log(\lambda L_\lambda),
\end{equation}
where $\lambda=510\,\si{nm}$ is the observed wavelength and $L_\text{iso}$ is the isotropic luminosity of
the quasar, which is taken to be equal to the bolometric luminosity determined above.
For each redshift bin we generate a sample of
$N=1000$ quasars, compute the impact model for each combination of $\pr$ and $q$, 
and finally derive the resulting ratios of maximum luminosity and
maximum flux ratios for observed wavelength $\lambda = 510\,\si{nm}$.

\begin{table}
    \centering
    \caption{Fraction of impacts exceeding the simulated quasar bolometric
    luminosity in each redshift bin, for each
    mass ratio and impact distance $\pr$.}
    \label{tb:impactlum}
\begin{tabularx}{\columnwidth}{Xcccccc}
    \toprule
    Mass ratio& \multicolumn{2}{c}{$0.1$} & \multicolumn{2}{c}{$0.3$} & \multicolumn{2}{c}{$1.0$} \\
    $\pr$&  1 & 10 & 1 & 10 & 1 & 10 \\
    Redshift & & & \\
    \midrule
    \mbox{0.2--0.4}  &   0.1\% & 0.0\% &   6.4\% & 0.1\% &  63.1\% & 7.5\% \\ 
    \mbox{0.4--0.6}  &   0.0\% & 0.0\% &   1.4\% & 0.0\% &  56.8\% & 1.7\% \\ 
    \mbox{0.6--0.8}  &   0.0\% & 0.0\% &   0.0\% & 0.0\% &  33.7\% & 0.0\% \\ 
    \mbox{0.8--1.0}  &   0.0\% & 0.0\% &   0.0\% & 0.0\% &   2.8\% & 0.0\% \\ 
    \mbox{1.0--1.2}  &   0.0\% & 0.0\% &   0.0\% & 0.0\% &   0.3\% & 0.0\% \\ 
    \mbox{1.2--1.4}  &   0.0\% & 0.0\% &   0.0\% & 0.0\% &   0.0\% & 0.0\% \\ 
    \mbox{1.4--1.6}  &   0.0\% & 0.0\% &   0.0\% & 0.0\% &   0.4\% & 0.0\% \\ 
    \mbox{1.6--1.8}  &   0.0\% & 0.0\% &   0.0\% & 0.0\% &   3.8\% & 0.0\% \\ 
    \mbox{1.8--2.0}  &   0.0\% & 0.0\% &   0.0\% & 0.0\% &   3.4\% & 0.0\% \\ 
    \bottomrule
\end{tabularx}
\end{table}

\begin{table}
    \centering
    \caption{Fraction of impacts exceeding the simulated quasar 
        flux at $\lambda=510\,\si{nm}$ in each redshift bin, for each
    mass ratio and impact distance $\pr$.}
    \label{tb:impactflux}
\begin{tabularx}{\columnwidth}{Xcccccc}
    \toprule
    Mass ratio& \multicolumn{2}{c}{$0.1$} & \multicolumn{2}{c}{$0.3$} & \multicolumn{2}{c}{$1.0$} \\
    $\pr$&  1 & 10 & 1 & 10 & 1 & 10 \\
    Redshift & & & \\
    \midrule
    \mbox{0.2--0.4}  & 1.7\% & 15.2\% &  13.8\% & 65.8 \% &  53.2\% & 97.1\% \\
    \mbox{0.4--0.6}  & 0.5\% &  2.5\% &  13.6\% & 56.1 \% &  59.9\% & 98.0\% \\
    \mbox{0.6--0.8}  & 0.1\% &  0.0\% &   5.7\% & 12.8 \% &  64.5\% & 98.7\% \\
    \mbox{0.8--1.0}  & 0.0\% &  0.0\% &   0.2\% &  0.0 \% &  53.3\% & 97.4\% \\
    \mbox{1.0--1.2}  & 0.0\% &  0.0\% &   0.1\% &  0.0 \% &  53.9\% & 55.5\% \\
    \mbox{1.2--1.4}  & 0.0\% &  0.0\% &   0.0\% &  0.0 \% &  48.5\% &  1.5\% \\
    \mbox{1.4--1.6}  & 0.0\% &  0.0\% &   0.4\% &  0.0 \% &  73.7\% &  2.9\% \\
    \mbox{1.6--1.8}  & 0.0\% &  0.0\% &   2.7\% &  0.0 \% &  71.4\% &  6.9\% \\
    \mbox{1.8--2.0}  & 0.0\% &  0.0\% &   3.1\% &  0.0 \% &  65.5\% &  2.1\% \\
    \bottomrule
\end{tabularx}
\end{table}

\begin{figure*}
\includegraphics[width=\textwidth]{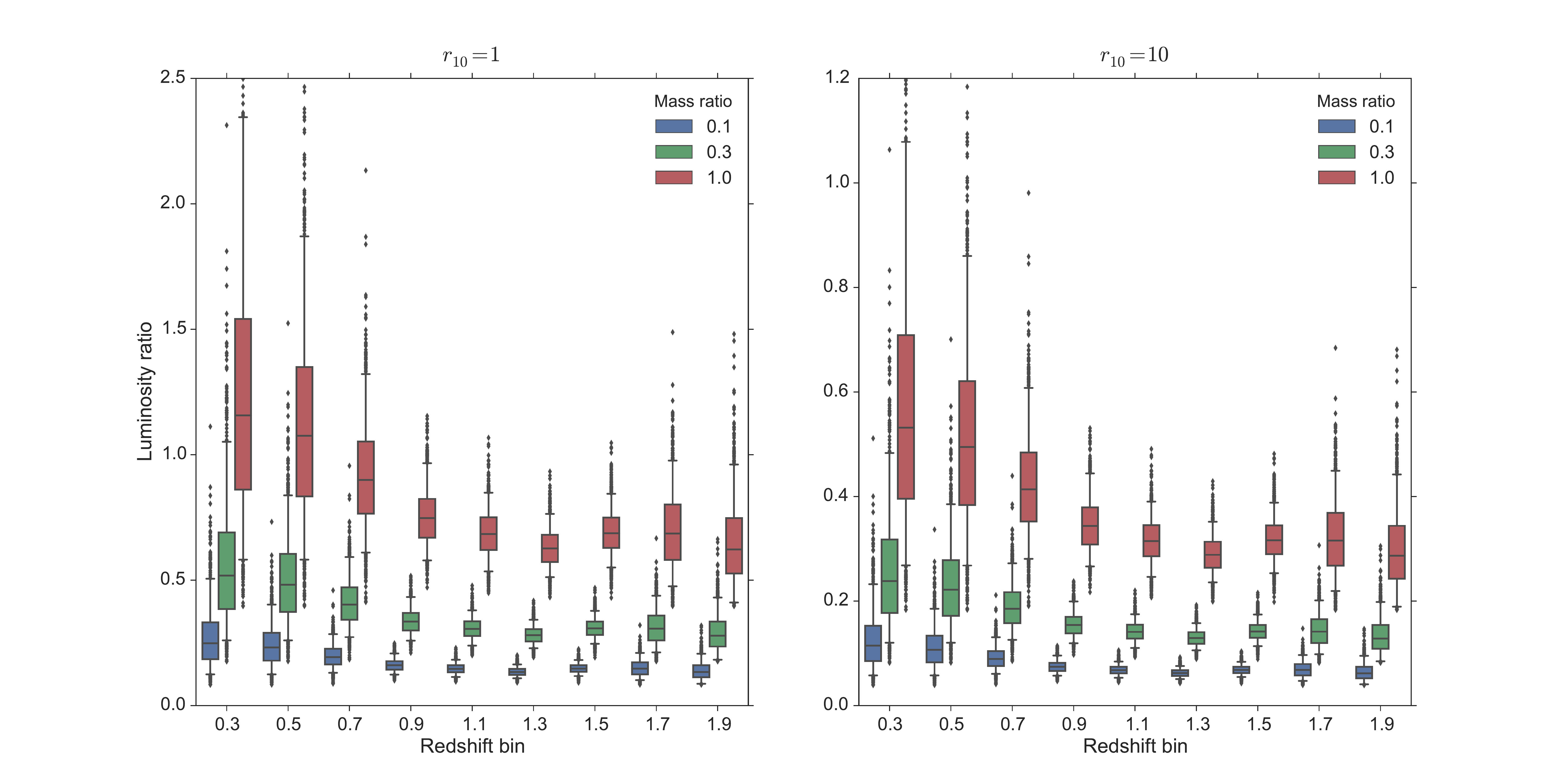}
\includegraphics[width=\textwidth]{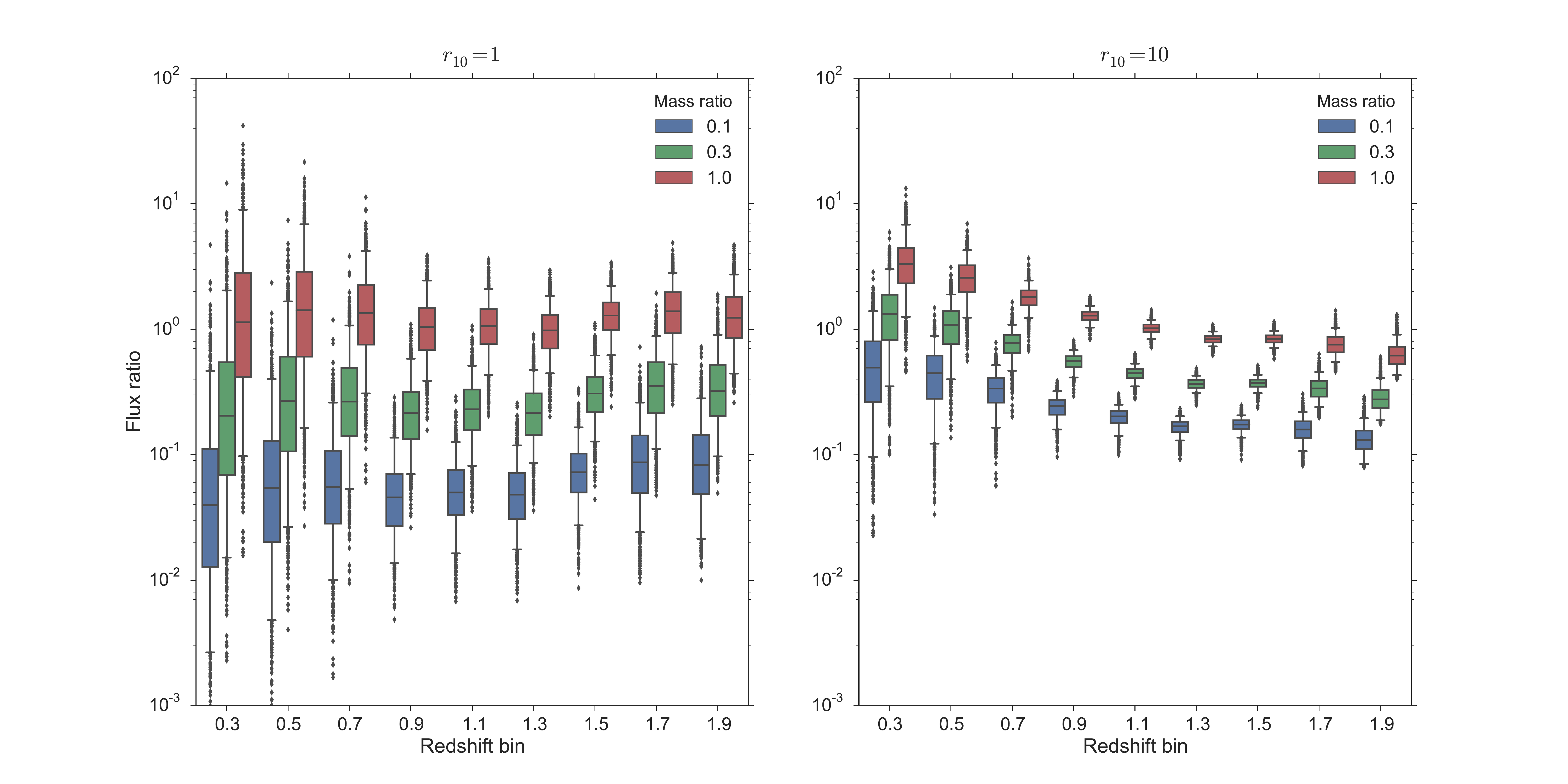}
\caption{\label{fig:ratioplot}
Top figure shows \rv{the} ratio of the maximum impact outburst bolometric luminosity to
the quasar bolometric luminosity as a function of redshift bin center
Bottom figure shows the ratio of the maximum outburst flux at
$\lambda=510\,\si{nm}$ to the quasar flux at the same observed wavelength.
The impacts are calculated for $\pr=1$ (left column) and
$\pr=10$ (right column) and mass ratios $q\in[0.1,0.3,1.0]$, indicated by blue, green
and red colours respectively. The boxes extend from first to third quartile,
with a line indicating the median value. The whiskers extend from 5th to 95th
percentile, with outlier values indicated by isolated points. Note the
logarithmic $y$-axis scaling in the lower plot.
}
\end{figure*}

The results are depicted graphically in Figure~\ref{fig:ratioplot}. 
In addition, the fraction of impact outbursts that equal or exceed the quasar
bolometric luminosity or flux at $\lambda=510\,\si{nm}$ are listed in
Tables \ref{tb:impactlum} and \ref{tb:impactflux}.
While the impact outburst
luminosity can exceed the quasar luminosity, this is only likely for
high mass ratios, small impact distances and lower redshifts.
The luminosity
ratio distribution does have a long tail with outliers with up to
a ratio of 5 for $\pr=1$ and $q=1.0$.
However, the observed flux at $\lambda=510\,\si{nm}$ from the impact outburst can
exceed the quasar flux by more than an order of magnitude.
For $q=1$, the median outburst flux is equal to the quasar flux up to
$z=1.8\text{--}2.0$, and up to $z=0.4\text{--}0.6$ for $q=0.3$.
\rv{From the above, we} conclude that combined with the expected spectral
evolution discussed in Section~\ref{sc:final_model}, it should be possible to
discern impact outbursts from the normal quasar stochastic variability using
observations in the optical frequencies.

\rv{\subsection{Observational complications and secondary effects}}

\rv{The physics of the interaction between a black hole and the
accretion disc present some complications to the idealized
observational picture presented above.
Furthermore, the sequence of impacts can have observable effects on a
larger (galactic) scale as well. We will briefly consider these in the
following.}

\rv{A central question is the effect of the secondary companion on the
of the accretion disc of the primary. If disc and binary co-align, or
the disc is disrupted, the sequence of observable discrete impacts will
cease.
In general, it is known that a circumbinary disc and the binary orbital
plane will eventually align \citep{ivanov1999,lodato2006,nixon2012b},
possibly through violent tearing of the disc \citep{nixon2012a,nixon2013}. However,
we are specifically interested in the case where a circumbinary cavity has not
yet formed. Precisely this situation was investigated in
\citet{ivanov1999}. They found that the accretion disc impacts would
continue until the disc interior to the binary aligns with the binary
orbital plane. They estimate the alignment to happen on a time-scale
\begin{equation}
    \frac{t_a}{\Porb} \sim \frac{M_2}{2\pi\Sigma \rbh^2 A}
    \sim 5.2\times
    10^{10}\,A^{-1}\alpha\mdot\pms^{-1}\pr^{-7/2}\incp^{4},
\end{equation}
where $t_a$ is the alignment time-scale, $\Porb$ is the binary orbital
period, $\Sigma$ is the accretion disc surface density,
$\mdot=\dot{M}/\Medd$, the constant $A>1$ depends on the disc response
on scales larger than $\rbh$, and we have used the inner, radiation
pressure dominated solution for $\Sigma$. This is valid for impacts up
to $\pr\sim 400\mdot^{16/21}(\alpha \pmp)^{2/21}$ \citep{shakura1973}.
We find that the alignment time is sufficiently long for extended series
of impact outbursts to occur.}

\rv{
However, the impacts also deplete the accretion disc of gas. This may
lead to a lower steady state surface density of the disc. This would
impact the time-scale of the outbursts, but the maximum luminosity (see
equation~\eqref{eq:Lmaxestimate}) would not be much affected. This is
because the lower energy density is compensated by the shortened
diffusion time-scale. The peak flux densities at a given wavelength
would be expected  to decrease, however.
Nevertheless, if the
depletion is fast enough, a cavity will be excavated and the outbursts
will cease. \citet{ivanov1999} give an approximate condition for
negligible depletion in terms of the mass ratio $q$,
\begin{equation}
    q < \sqrt{\frac{3\pi}{4}} \alpha^{1/2}\frac{h}{r} \sim
    1.7\,\alpha^{1/2}\mdot\pr^{-1}(1-0.55\pr^{-1/2}),
\end{equation}
where $h/r$ is the disc opening angle. We see that e.g.\ for the high
luminosity quasars ($\mdot\sim1$) considered above, the disc
accretion rate is sufficient to maintain the theoretical $\alpha$-disc
surface density at $\pr=10$ for up to $q\sim 0.04$. For higher mass
ratios, when the condition is violated,
the steady state surface density can be expected to decrease to a value 
\citep{ivanov1999}
\begin{equation}
    \Sigma = \alpha q^{-2} \left(\frac{h}{r}\right)^2\Sigma_0 \sim 
    1.2\,\alpha q^{-2} \mdot^2 \pr^{-2}(1-1.1\pr^{-1/2}),
\end{equation}
where $\Sigma_0$ is the unperturbed steady state surface density.
For $\alpha=0.1$, $\pr=10$ and $\mdot\sim 1$ we find $\Sigma \sim
8\times 10^{-4}\,q^{-2}\Sigma_0$. This is reasonably small, even for
mass ratios up to $q=0.1$. We may conclude that a train of accretion
disc impact outbursts can continue for a significant amount of time,
if the binary mass ratio is modest ($q\lesssim 0.1$) and the impacts happen
near the inner regions of the accretion disc of the primary. For larger
mass ratios, we expect the impact sequence to diminish in brightness
with a characteristic minimum time-scale of approximately
\begin{equation}
    t_\text{dim} \sim 5.3\times 10^{11}\,
    \mdot^{-2/3}\pmp^{7/3}\pms^{-4/3}\pr^{11/3}\incp^{8/3}\,\si{s},
\end{equation}
derived by dividing the accretion disc mass inside the binary orbit by
the approximate rate of matter expulsion by the binary,
$\sim2\pi (R_0^\text{LV})^2\Sigma_0/\Porb$. From this, we can estimate the number of
observable outbursts before significant reduction in the disc surface
density. We find
\begin{equation}
    \frac{2t_\text{dim}}{P}\sim 5.4\times 10^4\,
    \mdot^{-2/3}\pmp^{4/3}\pms^{-4/3}\pr^{13/6}\incp^{8/3},
\end{equation}
and conclude that even for $q=1$ the estimate yields around $10^2$
outbursts before the inner disc is depleted. However, as a final remark,
we caution that for $q\sim 1$, the secondary may violently disrupt the
inner disc on time-scales that are faster than the estimates above.}

\rv{
A related complication is that as the impacts modulate the structure of
the accretion disc, they can affect the accretion rate and thus
the luminosity of the primary SMBH. This luminosity change would be
apparent approximately after one sound-crossing time of disc,
\begin{equation}
    t_\text{sc} = \frac{r}{c_s} \sim \frac{r^2}{h\vkep} 
    \sim 2.9\times 10^6 \mdot^{-1} \pmp \pr^{5/2}\,\si{s}.
\end{equation}
Typically $t_\text{sc}\lesssim \Porb$, and observations of the impact
sequence will be complicated by the changing accretion rate of the
primary. An estimate for the change in the primary accretion rate 
can be obtained from the results of \citet{kumar2010}, who considered
the effect of supernova explosions in an accretion disc. Using their
results yields a mean increase in the primary accretion rate of
\begin{equation}
    \Delta\dot{M} \sim 4\times 10^{-3}\, q^{2.2}\pmp^{-0.25}
    \ph^{1.1}\pn^{0.85}\pr^{-0.075}\incp^{-2.2}\,\Medd.
\end{equation}
The corresponding change in luminosity is small compared to the impact
luminosity, which can be $\sim q\Ledd$. The change in primary black hole
luminosity also decreases faster with diminishing mass ratio. We thus
expect that the impact outbursts would not be completely overshadowed
by the changes in the primary black hole luminosity.
}

\rv{
Finally, there is the possibility that the impact may eventually cause a
wind-like outflow to form. This is dependent on the velocity to which
the outflow is accelerated. The impact velocity is $\vrel =
\sqrt{2}\incp\vkep$, and by the discussion in
Section~\ref{sc:disc_impact} we expect the outflow to reach velocities
higher than this. For high-inclination impacts, this is then above the
escape velocity. In this case, the outflow will eventually turn over
into a Sedov-Taylor blastwave. At the time when the radiative losses
become important, the outflow has reached a distance of
\begin{equation}
    R_\text{rad}\sim 210\,\pms^{10/17}
    \pn^{5/17}\pr^{5/17}\incp^{-10/17}n_0^{-1/5}\,\si{pc}
\end{equation}
at a velocity of 
\begin{equation}
    v_\text{rad}\sim 290 \,\pms^{2/17}
    \pn^{1/17}\pr^{1/17}\incp^{-2/17}n_0^{-1/5}\,\si{km.s^{-1}},
\end{equation}
where $n_0=n/(1\,\si{cm^{-3}})$ is the gas density in the galaxy nucleus.
Since the estimates
above are lower limits, and likely $n_0\lesssim 0.1$ \citep{mathews2003}, we see that the impact sequence can create a
wind at $\sim\si{kpc}$ scales, moving at velocities of $\sim 500\,\si{km.s^{-1}}$.
}

\rv{We can conclude that particularly binaries with mass ratios
$\lesssim 0.05$ can yield impact flare sequences that last for very long
times, and thus should in principle be observable. Furthermore,
evidence for a history of accretion disc impacts in galactic nucleus can
possibly be imprinted in winds originating from the nucleus. To complete
the discussion, we examine the assumptions made in building the analytic
model of the disc impact scenario itself in the next section.}

\section{Discussion}\label{sc:discussion}

The physical situation during an accretion disc impact in a supermassive
binary black hole system, such as OJ287, is obviously more complex
than what is assumed in LV96, I98 or this work. However, little analytic
progress can be made unless several simplifying assumptions are made.
Explicitly listed, the significant
assumptions shared by all these models are:
\begin{enumerate}
    \item spherical symmetry
    \item adiabatic expansion into empty space
    \item constant temperature profile
    \item constant density profile
    \item gravitational effects are negligible
    \item negligible magnetic fields
    \item negligible radiation from the accretion disc.
\end{enumerate}
We will discuss the possible effects these assumptions may have on the
validity of the model results.

The hydrodynamical simulations
in I98 show that spherical symmetry is actually a rather good
approximation for the problem. A steep density gradient in the
disc can be expected to produce a more conical outflow, but this
conicality will be partly masked by the relativistic restrictions on
the visible angular extent of the photosphere.
However, I98 also find that the density
profile of the outflow sphere is given by a steep power law, $\rho\propto
R^{-5}$. If this is true, the outer layers would be less dense than in the
homogeneous case, decreasing their optical thickness.
This decrease is partly balanced by the fact the outer layers should
also be somewhat cooler, both initially and during the evolution. This will tend
to increase the opacity when temperatures are in the region dominated by
Kramers' law. In this sense, the compromise of homogeneous density and
temperature profiles is acceptable.

The gravitational effects of the accretion disc
and the primary on the evolution of the outflow gas
are difficult to quantify. However,
at least for impacts happening close to the primary,
neglecting gravitational effects is likely not a good
approximation, whereas the gravitational perturbation caused by the accretion
disc is negligible compared to the black holes. Strong gravity near the
primary would certainly distort the outflow into an asymmetric shape.
This geometric distortion might affect both peak luminosity and outburst
time-scale due to decreased diffusion time and earlier onset of optical
thinness. More importantly, as investigated in e.g. \citet{karas1994} and \citet{dai2010},
the gravitational lensing effect might increase the observed flux by a
large factor. The observed spectrum would also be affected by
the differences in light travel times from different parts of the
outflow, as well as gravitational redshift.

Observations and models suggest that non-negligible magnetic fields
exist near the
inner parts of the accretion disc and threading the central black hole in
luminous AGN
\citep{eatough2013,sikora2013,silantev2013,zamaninasab2014,martividal2015}.
Unfortunately, the expected magnetic field strengths are not well
constrained, and the degree of ordering is likewise uncertain
\citep{king2007,mckinney2007,tchekhovskoy2011}.
Nevertheless, even a disordered
magnetic field would still produce a synchrotron component in the
observed radiation. The synchrotron emission would serve to convert some
of the plasma kinetic energy to radiation, likely increasing the outburst
duration. However, in the case of OJ287, there is evidence that outbursts
associated with accretion disc impacts are associated with lower degree of
polarization \citep{val2008}. This indicates that synchrotron emission is not a
major component during these outbursts.

The accretion disc serves not only as the source of the outflow
material, but it also illuminates it and reflects the outflow radiation
back. These effects are difficult to estimate analytically, though we
may assume that they are significant especially near the hot inner
regions of the accretion disc.

Lastly, the assumption of expansion into free space is not
unproblematic either, since it is commonly assumed that accretion discs
are embedded in hot coronae. This corona of hot rarified gas might
comptonize the radiation from the outflow. Furthermore, while an
accretion disc corona is usually expected to be rarified, this might not
be the case if recurrent impacts have separated a significant amount of
matter from the accretion disc to the corona. This
matter might then serve to both cloud the outflow from view, and
decelerate the outflow, turning a linear expansion into a Sedov--Taylor
power law.

The discussion above may cast doubt on whether any analytic model can be
expected to describe the accretion disc impact outflows to any degree of
accuracy. To make a case that this is possible, we should first note that 
if the impact site is distant from the primary black hole, all of the
assumptions with the exception of (iii) and (iv) are valid. And, as
already noted, even if assumptions (iii) and (iv) are violated, this should not
greatly affect the luminosity or the spectral evolution, at least initially. 
When the impact site 
is taken to be progressively closer to the primary, the assumptions will break
down, but only gradually, to some order of $\pr$. Even near the primary
Schwarzschild radius at $\pr=0.1$ we may expect the estimated luminosity to be
of the correct order, even though the evolution of observed flux may have
diverged completely from the model estimates. In conclusion, despite the caveats
mentioned above, we expect the model to have good predictive power in much of
the SMBH binary orbital parameter space.

\section{Conclusions}\label{sc:conclusions}

We have estimated the physical characteristics of an impact by a supermassive
black hole on the accretion disc of another supermassive black hole. We have
compared our results to earlier studies, with a focus on the LV96 paper, and we
find that an impact would cause an expanding outflow to form. This outflow
expands at a mildly relativistic velocity, and is observable very soon after the
initial impact, with most of the radiation escaping through diffusion.

Building on these estimates, we have developed a model for the spectral evolution
of such an impact outflow. The model is based on time-dependent photon diffusion
and takes into account the emission from optically thin regions. The observed
flux is then corrected for relativistic effects. In addition to
the model, we present approximate analytic formulae for estimating maximum
luminosity, maximum flux and the relevant time-scales, correct to first order 
in outflow expansion velocity $\beta$.

The model was applied to simulated populations of quasars in a redshift range
$z=0.2\text{--}2.0$, constructed by using observed quasar statistics. The
quasars were assumed to host SMBH binaries with a selection of mass ratios and
orbital radii, and the accretion disc parameters of the primary black hole were
estimated with thin $\alpha$-disc theory.
The maximum luminosity and the flux at the observed wavelength $\lambda=510\,\si{nm}$
resulting from the accretion disc impact model for each configuration
were computed.
The model results were compared to the quasar bolometric luminosity and flux at
the same observed wavelength.
The quasar luminosities were set to the sub-Eddington limit (see
\citet{steinhardt2010} and Section~\ref{sc:observational_tool}), and the
quasar fluxes were derived using the bolometric corrections in
\citet{runnoe2012}.
The results indicate that
detecting impact flares should be possible, as the impact flares can peak at flux
densities that are several times the value of the quasar flux, and can
be comparable up to high redshifts with moderate binary mass ratios.
As such, searching
for impact outbursts might serve as an independent method for locating SMBH
binaries in active galactic nuclei.

Finally, we have extensively discussed the simplifying assumptions that have
been made in producing the model and the analytic results for the impact
outflows. While most of the assumptions are violated for impacts happening near
the primary black hole, we expect that in most cases our model should give a
good description of the outburst.  However, it should be emphasized that the
problem would greatly benefit from a comprehensive assessment with
state of the art numerical simulations.

\section*{Acknowledgements}

\rv{
I wish to thank the referee, Tamara Bogdanovic, for pointing out some
oversights and sections in need of further elaboration. I believe these
changes have substantially improved the article.
I am also grateful to
Peter H.~Johansson for a multitude of
helpful comments on the original manuscript, and
to Mauri Valtonen and Joonas Nättilä for fruitful
discussions on the subject.
Finally, I am indebted to Harry Lehto for the
invaluable access to the original research material of the LV96
publication.}
This research was supported by the Emil Aaltonen
foundation, and the Academy of Finland grants 1274931 and 267040.

%

\bibliography{mn-jour,references}

%
\appendix

\section{Symbol list}\label{sc:symbol_list}

Table~\ref{tb:symboltable} lists most of the mathematical symbols used
in the paper with explanations. The table is divided into sections that
roughly correspond to the sections of the paper. Within the sections,
the symbols are listed in the order of appearance, where feasible.

\begin{table*}
    \centering
    \caption{A summary of the symbols used in the paper.}
    \label{tb:symboltable}
    \begin{tabularx}{\textwidth}{lXlX}
        \toprule
        Symbol & Description & Symbol & Description  \\
        \midrule

        \multicolumn{4}{c}{\emph{Black hole binary \& accretion disc}}\\
        $\pmp$   & Primary black hole mass / $10^{10}\Msun$          & $\ph$       & Accretion disc semiheight / $10^{15}\si{cm}$ \\
        $\pms$   & Secondary black hole mass / $10^{8}\Msun$         & $\pn$       & Accretion disc number density / $10^{14}\,\si{cm^{-3}}$ \\
        $\rschp$ & Primary black hole Schwarzschild radius           & $\plumdist$ & Luminosity distance  / $1500\,\si{Mpc}$\\
        $\pr$    & Distance from the primary black hole / $10\rschp$ & $z$         & Redshift \\

        \multicolumn{4}{c}{\emph{Impact shock}}\\
        $k$        & Boltzmann constant                                                        & $n_2$                 & Post-shock number density \\
        $\sigma_T$ & Thomson electron scattering cross section                                 & $x$                  & Gas compression factor \\
        $\kappa_T$ & Thomson electron scattering opacity                                       & $\gamma_a$           & Adiabatic constant \\
        $\sigma_B$ & Stefan-Boltzmann constant                                                 & $\gamma$             & Electron Lorentz gamma factor \\
        $a$        & Radiation constant, $4\sigma_B/c$                                         & $\beta$              & Lorentz beta \\
        $m_p$      & Proton mass                                                               & $t_\text{ff}$        & Bremsstrahlung cooling time-scale \\
        $m_e$      & Electron mass                                                             & $t_\text{IC}$        & Inverse Compton cooling time-scale \\
        $\vkep$    & Keplerian orbital velocity of the impactor                                & $\epsff_{\nu}$       & Bremsstrahlung volume emissivity \\
        $\vrel$    & Relative velocity of impactor and accretion disc gas                      & $\epsff$             & Integrated bremsstrahlung volume emissivity \\
        $\incl$    & Inclination angle between the binary orbital plane and the accretion disc & $\epsilon_\text{IC}$ & Inverse Compton volume emissivity \\
        $\incp$    & Inclination parameter, $\sqrt{1-\sin\incl}$                               & $\Teq$               & Shock equilibrium temperature \\
        $c_s$      & Speed of sound in the accretion disc                                      & $\mach_1$            & Shock mach number \\
        $\tdyn$    & Dynamical time-scale of the impact event                                  & $\mach_{e,1}$        & Shock effective mach number \\
        $\Tdisc$   & Accretion disc temperature                                                & $\Pgas$              & Gas pressure \\
        $\Tvir$    & Accretion disc temperature                                                & $\Prad$              & Radiation pressure \\
        $n$        & Pre-shock number density                                                  & $R_P$                & Ratio of radiation to gas pressure \\
        
        \multicolumn{4}{c}{\emph{Outflow}}\\
        $R_0^\text{I98}$  & Initial outflow radius according to I98               & $V_0$     & Initial outflow volume \\
        $R_0^\text{LV96}$ & Initial outflow radius according to LV96              & $\lambda$ & Ratio of secondary Bondi--Hoyle radius to accretion disc semiheight \\
        $R_0^\text{sim}$  & Initial outflow radius estimated from I98 simulations & $\Ledd$   & Eddington luminosity \\
        $\rbh$            & Bondi--Hoyle radius of the secondary black hole       & $\tdiff$  & Photon diffusion time-scale\\

        \multicolumn{4}{c}{\emph{LV96 model}}\\
        $P$          & Total pressure                                               & $\expfacff$ & Expansion factor when $\taua = 1$ \\
        $\rho$       & Gas density                                                & $\expfacT$  & Expansion factor when $\taus = 1$ \\
        $R$          & Radial size of the outflow bubble                          & $t_0$       & Time delay between the impact and the outflow turning optically thin \\
        $T$          & Temperature of the outflow bubble                          & $S_V$       & Flux density in the optical $V$-band \\
        $u_1$, $u_2$ & Pre- and post-shock velocities of the gas in shock frame   & $\toutb$    & Length of the outburst \\
        $v_2$        & Post-shock gas velocity in observer frame                  & $\ffcutoff$ & Bremstrahlung volume emissivity exponential cutoff parameter \\
        $\expfac(t)$ & Relative expansion factor $R(t)/R_0$                       & $W_{-1}$    & The $-1$ branch of the Lambert W function \\
        $\mu$        & Mean molecular weight                                      & $T_e$       & Electron temperature \\
        $\thyd$      & Hydrodynamical, or expansion time-scale                    & $y$         & The comptonization parameter \\
        $\taueff$    & Effective optical depth                                    & $\Teff$     & Effective black body temperature\\
        $\taua$      & Optical depth of absorption processes                      & $\dilution$ & Dilution factor of the diluted black body spectrum\\
        $\taus$      & Electron scattering optical depth                          & $E_r$       & Radiative energy \\
        $\kappa_a$   & Kramers' law of opacity                                    & $\eta$      & Ratio of radiative energy to matter rest energy \\
        $K_a$        & The constant factor of Kramers' law                        & $\Lthin$    & Optically thin luminosity \\
        $\expfacLV$  & Expansion factor when $\sqrt{\taua\taus} = 1$              & $\Ldiff$    & Optically thick, diffusive luminosity \\

        \multicolumn{4}{c}{\emph{New model}} \\
        $\kappa$     & Opacity                                 &  $\Ltot$          & Total luminosity                           \\
        $x$          & Relative radial distance, $x=r/R(t)$    &  $l$ & Photon mean free path                                   \\
        $\psi(x)$    & Dimensionless temperature profile       &  $\tau_p$ & Optical depth of the photosphere                   \\
        $\phi(t)$    & Thermal energy time dependency          &  $x_\sigma$ & Relative radial distance at which $\taus=\tau_p$ \\
        $\alpha$     & Diffusion equation eigenvalue           &  $x_e$ & Relative radial distance at which $\taueff=\tau_p$    \\
        $\eta(x)$    & Dimensionless density profile           &  $x_p$ & Relative radial distance of the photosphere           \\
        $M(x)$       & Total mass within $x$                   &  $t_c$ & Generic cooling time-scale                            \\
        $I_M(x)$     & Density distribution form factor        &  $\beta$ & Expansion velocity / $c$                            \\
        $E_T(x,t)$   & Thermal energy within $x$ at time $t$   &  $\Gamma$ & Expansion Lorentz gamma factor                     \\
        $I_T(x)$     & Thermal energy distribution form factor &  $t_e$ & Emission time                                         \\
        \bottomrule
    \end{tabularx}
\end{table*}

\section{Code}\label{sc:code}

A Python/Cython\footnote{%
See \url{http://www.python.org} and \url{http://cython.org}.%
}
computer code was constructed to perform the model computations used in
the paper. This code is freely available at
\url{https://bitbucket.org/popiha/impact_model}. The code can be used to
compute the accretion disc impact spectrum at given 
frequencies, as well as other model quantities as a function of time.
Detailed information and instructions for use
are provided with the code.

\bsp
\label{lastpage}
\end{document}